\documentclass{aa}  

\usepackage{graphicx}
\usepackage{txfonts}
\usepackage{soul,xcolor}
\setstcolor{red}
\usepackage[colorlinks=true, allcolors=blue]{hyperref}
\begin{document}

   \title{EWOCS-III: JWST observations of the supermassive star cluster Westerlund 1}

   \subtitle{}

   \author{M. G. Guarcello\inst{1}\thanks{mario.guarcello@inaf.it}
          \and
          V. Almendros-Abad\inst{1}
          \and
          J. B. Lovell\inst{2}
          \and
          K. Monsch\inst{2}
          \and
          K. Mu\v{z}i\'c\inst{3}
          \and
          J. R. Mart\'inez-Galarza\inst{2}
          \and
          J. J. Drake\inst{4}  
          \and
          K. Anastasopoulou\inst{2}
          \and          
          M. Andersen\inst{5} 
          \and
          C. Argiroffi\inst{6,1}          
            \and
          A. Bayo\inst{5}
            \and
          R. Bonito\inst{1}
           \and
          D. Capela\inst{3} 
           \and
          F. Damiani\inst{1}
           \and
          M. Gennaro\inst{7}   
           \and
          A. Ginsburg\inst{8} 
            \and
          E. K. Grebel\inst{9} 
          \and
          J. L. Hora\inst{2}
          \and
          E. Moraux\inst{10} 
           \and
          F. Najarro\inst{11}
          \and
          I. Negueruela\inst{12} 
          \and
          L. Prisinzano\inst{1}
          \and
          N. D. Richardson\inst{13} 
          \and
          B. Ritchie\inst{14} 
          \and
          M. Robberto\inst{7} 
          \and
          T. Rom\inst{15,10} 
          \and
          E. Sabbi\inst{16,7,17}   
          \and
          S. Sciortino\inst{1}
          \and
          G. Umana\inst{18} 
          \and
          A. Winter\inst{19,10} 
          \and
          N. J. Wright\inst{20} 
           \and
          P. Zeidler\inst{21} 
          }

   \institute{
        Istituto Nazionale di Astrofisica (INAF) -- Osservatorio Astronomico di Palermo, Piazza del Parlamento 1, 90134 Palermo, Italy\\  
        \email{mario.guarcello@inaf.it}
    \and 
        Center for Astrophysics $\vert$ Harvard \& Smithsonian, 60 Garden Street, Cambridge, MA 02138, USA
   \and 
        Instituto de Astrofísica e Ciências do Espaço, Faculdade de Ciências, Universidade de Lisboa, Ed. C8, Campo Grande, 1749-016 Lisbon, Portugal             
   \and 
        Lockheed Martin, 3251 Hanover St, Palo Alto, CA 94304
   \and 
        European Southern Observatory, Karl-Schwarzschild-Strasse 2, D-85748 Garching bei M\"{u}nchen, Germany
   \and 
        Department of Physics and Chemistry, University of Palermo, Piazza del Parlamento 1, 90134 Palermo, Italy
   \and 
        Space Telescope Science Institute, 3700 San Martin Dr, Baltimore, MD, 21218, USA
   \and 
        Department of Astronomy, University of Florida, P.O. Box 112055, Gainesville, FL 32611-2055, USA
   \and 
        Astronomisches Rechen-Institut, Zentrum f\"ur Astronomie der Universit\"at Heidelberg, M\"onchhofstr.\ 12--14, 69120 Heidelberg
   \and 
        Univ. Grenoble Alpes, CNRS, IPAG, 38000 Grenoble, France
   \and 
        Departamento de Astrofísica, Centro de Astrobiología, (CSIC-INTA), Ctra. Torrejón a Ajalvir, km 4, Torrejón de Ardoz, E-28850 Madrid, Spain
   \and 
        Departamento de Física Aplicada, Facultad de Ciencias, Universidad de Alicante, Carretera de San Vicente s/n, E-03690, San Vicente del Raspeig, Spain
   \and 
        Department of Physics and Astronomy, Embry-Riddle Aeronautical University, 3700 Willow Creek Rd., Prescott, AZ 86301, USA
   \and 
        School of Physical Sciences, The Open University, Walton Hall, Milton Keynes MK7 6AA, UK
   \and 
        University of Split, Faculty of Science, Department of Physics, Ruđera Boškovića 33, 21000, Split, Croatia
   \and 
        Gemini Observatory/NSF’s NOIRLab, 950 North Cherry Avenue, Tucson, AZ, 85719, USA
   \and 
        Steward Observatory, University of Arizona, 933 North Cherry Avenue, Tucson, AZ 85721, USA
   \and 
        Istituto Nazionale di Astrofisica (INAF) – Osservatorio Astrofisico di Catania, Via Santa Sofia 78, I-95123 Catania, Italy
   \and 
        Universit\'{e} C\^{o}te d'Azur, Observatoire de la C\^{o}te d'Azur, CNRS, Laboratoire Lagrange, F-06300 Nice, France
   \and 
        Astrophysics Group, Keele University, Keele, Staffordshire ST5 5BG, United Kingdom
   \and 
        AURA for the European Space Agency (ESA), ESA Office, Space Telescope Science Institute, 3700 San Martin Drive, Baltimore, MD 21218, USA
}
   \date{}
  \abstract
   {The typically large distances, extinction, and crowding of Galactic supermassive star clusters (stellar clusters more massive than 10$^4\,$M$_\odot$) have so far hampered the identification of their very low mass members, required to extend our understanding of star and planet formation, and early stellar evolution, to the extremely energetic star-forming environment typical of starbursts. This situation has now evolved thanks to the \textit{James Webb} Space Telescope (JWST), and its unmatched resolution and sensitivity in the infrared.}
   {In this paper, the third of the series of the Extended Westerlund 1 and 2 Open Clusters Survey (EWOCS), we present JWST/NIRCam and JWST/MIRI observations of the supermassive star cluster Westerlund 1. These observations are specifically designed to unveil the cluster members down to the brown dwarf mass regime, and to allow us to select and study the protoplanetary disks in the cluster and to study the mutual feedback between the cluster members and the surrounding environment.}
   {Westerlund~1 was observed as part of JWST GO-1905 for 23.6 hours. The data have been reduced using the JWST calibration pipeline, together with specific tools necessary to remove artifacts, such as the $1/f$ random noise in NIRCam images. Source identification and photometry were performed with \textit{DOLPHOT}.}
   {The MIRI images show a plethora of different features. Diffuse nebular emission is observed around the cluster, which is typically composed of myriads of droplet-like features pointing toward the cluster center or the group of massive stars surrounding the Wolf-Rayet star W72/A. A long pillar is also observed in the northwest. The MIRI images also show resolved shells and outflows surrounding the M-type supergiants W20, W26, W75, and W237, the sgB[e] star W9 and the yellow hypergiant W4. Some of these shells have been observed before at other wavelengths, but never with the level of detail provided by JWST. The color-magnitude diagrams built using the NIRCam photometry show a clear cluster sequence, which is marked in its upper part by the 1828 NIRCam stars with X-ray counterparts. NIRCam observations using the F115W filter have reached the 23.8 mag limit with 50\% completeness (roughly corresponding to a 0.06$\,$M$_\odot$ brown dwarf).}
  {}
  \maketitle

\section{Introduction} \label{sec:intro}

Young stellar clusters with masses exceeding 10$^4\,$M$_\odot$ are typically referred to as supermassive star clusters \citep[SSCs; ][]{PortegiesZwart2010ARAA..48..431P}. These clusters serve as laboratories with which to expand our understanding of star and planet formation, as well as early stellar evolution, in the stellar environments characterized by rich and dense ensembles of very massive stars, which are typically found at large distances from the Sun. Despite being rare in the Milky Way today, such massive star-forming environments are common in galaxies undergoing starburst episodes, such as interacting galaxies \citep[e.g.,][]{LarsonTinsley1978ApJ...219...46L,Smith2010AJ....140.1975S}. \par

In the Milky Way, the sample of supermassive star clusters younger than 10$\,$Myr (which may still harbor protoplanetary disks and active star formation) contains fewer than ten members. All of these are situated at distances greater than 2.5$\,$kpc and are affected by large visual extinctions \citep{PortegiesZwart2010ARAA..48..431P}. High stellar densities, substantial extinction, and the large distances have hindered attempts to analyze their low-mass stellar populations, which are essential in order to understand star and planet formation in such massive environments. \par

In its first three years of operations, the \textit{James Webb} Space Telescope \citep[JWST;][]{Gardner2006SSRv..123..485G} has already proven to be a game-changing mission in this field. With its unparalleled performance in the infrared, JWST offers the nivel opportunity to unveil the low-mass stellar population of supermassive star clusters in the local Universe, study the environments around their most massive stars, and explore the properties of the parental clouds where these are still present. For example, the JWST Near InfraRed Camera \citep[NIRCam,][]{Rieke2023PASP..135b8001R} has allowed the selection and study of low-mass young stellar objects (YSOs) in 30~Doradus, the most massive star-forming region known in the local Universe \citep{Fahrion2023AA...671L..14F,Fahrion2024AA...681A..20F}, in the massive young cluster NGC~346 in the Small Magellanic Cloud \citep{Jones2023NatAs...7..694J,Habel2024ApJ...971..108H}, and even in the low-metallicity dwarf galaxy NGC~6822, located at 490$\,$kpc \citep{Lenkic2024ApJ...967..110L,Nally2024MNRAS.531..183N}. Spectroscopic observations with the JWST Mid-Infrared Instrument (MIRI) in the Medium Resolution Spectroscopy mode \citep{Rieke2015PASP..127..665R,Wright2023PASP..135d8003W} have allowed the measurement of accretion rates and, in one case (the star Y3, approximately 10 Myr old), the detection of prominent CH$_4$, NH$_3$, CH$_3$OH, CH$_3$OCHO, and CO$_2$ ice absorption features in 11 YSOs in the massive star-forming regions of N~79, in the Large Magellanic Cloud \citep{Nayak2024ApJ...963...94N}. \par

Similarly, detecting brown dwarfs (BDs) in massive star clusters was a challenging task in the pre-JWST era due to the inherent faintness of such objects, crowding, and the contamination from foreground and background sources. This in particular hindered our ability to understand whether or not the low-mass end of the initial mass function (IMF) and the production of BDs in starburst can be affected by their surrounding environment. Several studies have showcased the capability of JWST to detect BDs, enabling photometric and spectroscopic characterization of their populations and properties. For instance, JWST has been used to identify BD candidates not only in nearby star-forming regions, such as IC~348 in the Perseus molecular cloud, at a distance of $d\simeq 325\,$pc \citep{Luhman2024AJ....167...19L}, the Trapezium in Orion \citep[$d\simeq400\,$pc,][]{Pearson2023arXiv231001231P}, and NGC 1333 \citep[300-350$\,$pc][]{Langeveld2024AJ....168..179L}, but also in the extremely crowded globular cluster 47~Tucanae  \citep[$d\simeq 4.5\,$kpc,][]{Marino2024ApJ...965..189M}. \par 

JWST observations of the massive stars in supermassive star clusters can shed light on how the dynamical and radiative feedback from these stars impact their surrounding environment and the overall star formation process within their parental clouds. In very massive clusters, where most of the massive stars form, the winds emitted by these stars are exposed to incident UV radiation from other surrounding massive stars, and wind--wind interactions are favored due to the typically high stellar density of these clusters. Such processes may facilitate the confinement and cooling of the ejected material in the potential well of the cluster, providing a mechanism that could enable the formation of a new generation of stars from this enriched material. This mechanism is proposed to explain the existence of multiple stellar populations observed in globular clusters \citep[e.g.,][]{Piotto2007ApJ...661L..53P}. \par

The unique feature of Westerlund~1 is its large, dense, and diverse population of massive stars, which has no counterpart in other known Milky Way clusters in terms of the number of stars and the richness of their spectral types and evolutionary phases \citep{Clark2005A&A...434..949C,Ritchie2009AA...507.1585R,Clark2020AA...635A.187C}. The cluster hosts 24 Wolf-Rayet (WR) stars, several blue and yellow hypergiants (BHG/YHG), four red supergiants (RSGs), and a luminous blue variable (LBV), along with over 100 bright OB supergiants dominated by spectral classes O9-B1 \citep{Negueruela2010}. JWST observations of these evolved massive stars can unveil valuable information about their circumstellar environment, which is influenced by their winds and episodic mass--loss events. These are fundamental processes in the evolution of very massive stars, in particular  for stars in the 10-40$\,$M$_\odot$ mass range, which, during their evolution, undergo the RSG and/or YHG phases, which are characterized by dramatic mass loss, both in the form of steady and dense winds, as well as more impulsive short-lived ejection events \citep[e.g.,][]{Shenoy2016AJ....151...51S}. \par

JWST has already proven to be an ideal telescope for studying dust-rich shells surrounding evolved massive stars, as demonstrated in cases such as WR137 \citep{Lau2024ApJ...963..127L} and the carbon-rich WR binary star WR140 \citep{Lau2022NatAs...6.1308L}. MIRI observations have identified 17 dusty shells in WR140, which appear to be the result of periodic events of dust formation with a period of 7.93 years occurring over the last 130 years. These observations confirm the significant role of WR binaries in enriching the interstellar medium with dust. \par

In this paper, we present JWST NIRCam and MIRI observations of the young superstar cluster Westerlund~1, the closest cluster of this class \citep[d=$4.23^{+0.23}_{-0.21}\,$kpc,][]{Negueruela2022AA...664A.146N}. The age and total mass of Westerlund 1 remain subjects of debate. Several authors, including \citet{Clark2005A&A...434..949C}, \citet{Brandner2008AA...478..137B}, \citet{Gennaro2011}, and \citet{Andersen2017AA.602A.22A}, have estimated the total mass to range from approximately 5$\times$10$^4$ M${_\odot}$ to over 10$^5$ M${_\odot}$, and an age of around 5$\,$Myr \citep[see also][]{Kudryavtseva2012ApJ.750L.44K}. This estimate would make Westerlund 1 the most massive known stellar cluster in the Milky Way. In contrast, a higher age of approximately 10$\,$Myr has been suggested by \citet{Beasor2021ApJ...912...16B} and \citet{Navarete2022MNRAS.516.1289N}, which would also imply a lower total mass for the cluster than previously estimated. \par

The JWST observations are part of the Extended Westerlund 1 and 2 Open Clusters Survey (EWOCS)\footnote{\url{https://Westerlund1survey.wordpress.com/}}, and are designed to allow selection  of the low--mass content, into the brown dwarf regime, as well as its circumstellar disk population. This  will allow us to finally assess whether the evolution of protoplanetary disks, the formation of brown dwarfs, and the IMF slope are affected by starburst environment. The project, as well as the list of X-ray sources in and around the cluster from a 1$\,$Msec Chandra/ACIS-I large program, are presented in \citet{Guarcello2024AA...682A..49G}. \par

The paper is organized as follows: The JWST observations and data reduction workflow are discussed in Sect. \ref{sec:redux}, while in Sect. \ref{sec:features} we provide a qualitative description that includes the possible origin of the main features in the NIRCam color--magnitude diagrams (Sect. \ref{sec:nircamphot}), the nebulosity surrounding the cluster (Sect. \ref{sec:nebulosity}), and the shells and outflows in the red supergiant stars (Sect. \ref{sec:shells}) observed by MIRI. We present our conclusions in Sect. \ref{sec:conclusions}

\section{Data reduction, calibration, and images} \label{sec:redux}

We observed Westerlund 1 with NIRCam using five wide and six narrow filters and with MIRI with three filters in program GO 1905 (PI: Guarcello). Table \ref{tab:MIRIobssetup} shows the observation dates and integration parameters for the NIRCam and MIRI observations of Westerlund~1 (SCI) as well as a control field (CF), which is necessary for decontamination of the cluster locus in the color--color and color--magnitude diagrams, and is located at 16:47:43.0 -46:03:47.0. The CF is the same as that adopted for the \textit{Hubble} Space Telescope (HST) observations of Westerlund~1\citep{Andersen2017AA.602A.22A}. With NIRCam we adopted the \textit{fullbox-2TIGHTGAPS} primary dither and STANDARD subpixel dithers, using a 1$\times$3 mosaic in order to cover a large portion of the field observed with \texttt{Chandra}/ACIS-I. With MIRI, we used a 3$\times$3 mosaic and adopted a 4-Points dither set.  The images of the control field are shown in the Appendix \ref{appendix:cf}. The control field was observed adopting a similar setup. 

\subsection{Data calibration steps}
\label{sec:datared}
We started the reduction from the level 2 products (\texttt{uncal.fits} files). We used the JWST calibration pipeline v. 1.14.0, Python version 3.12.2, Astropy \citep{Astropy2013A&A...558A..33A,Astropy2018AJ....156..123A,Astropy2022ApJ...935..167A} version 6.0.0 , Numpy version 1.25.2, and the calibration reference data system version 11.17.14 (jwst\_1215.pmap). The basic NIRCam and MIRI data--reduction workflow is divided into three stages, as described in \citet{Bushouse_JWST_Calibration_Pipeline_2024}. 


   \begin{figure*}
   \centering
   \includegraphics[width=1\textwidth]
   {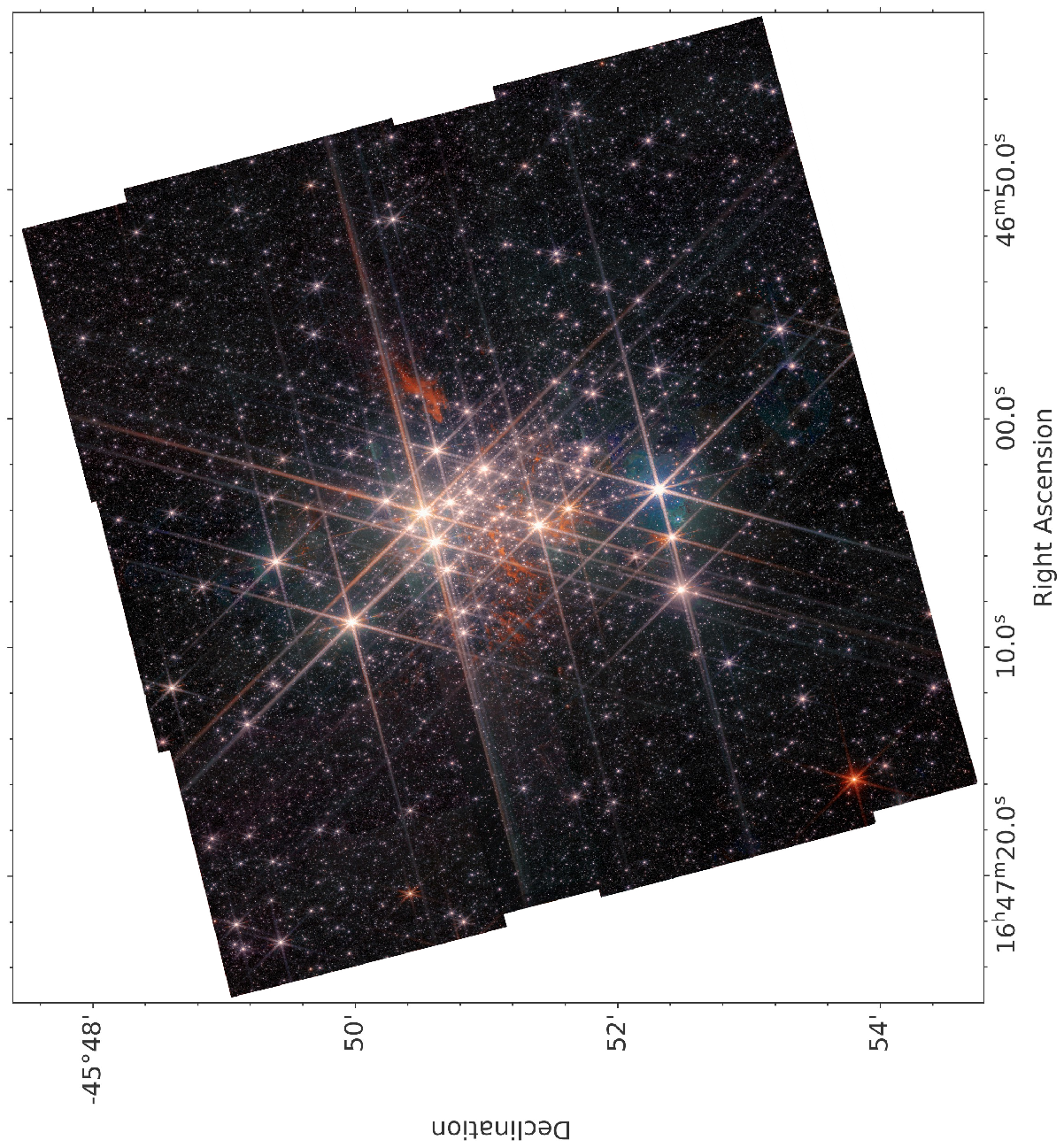}
      \caption{RGB NIRCam image of Westerlund 1 (red: F444W, green: F323N, blue: F277W). At 4230$\,$pc, the angular size of this image spans an area of 7.4$\times$7.4$\,$pc.}
         \label{Fig:nircam_mosaic}
   \end{figure*}

\subsubsection{NIRCam}
\label{sec:nircamred}
We removed the ``streaking'' features caused by the $1/f$ noise using a routine provided by C. Willott\footnote{https://github.com/chriswillott/jwst}. This method consists in the subtraction of the median of each background-subtracted row from each \texttt{cal.fits} exposure. We decided not to separate the correction of each exposure into the four individual amplifiers because the presence of strongly saturated stars hinders the correct calculation of the background and may introduce artifacts. 


While the mosaic of the control field was obtained by running the Stage 3 calibration in default mode, we found that the quality of the mosaic produced by the standard Stage 3 reduction of the Westerlund~1 field images was very poor, with strong misalignments. Some of the individual frames showed a significant offsets (up to several arcseconds) from others, an issue probably related to guiding. We find that the pipeline task \textit{tweakreg} does not deal correctly with these offsets when fed several frames at the same time, but performs well when run on a single file. Another important point for the alignment is the input astrometric reference catalog.
Initially, we used the catalog published by \citet{Andersen2017AA.602A.22A}, which is based on HST observations, combined with the \textit{Gaia} Third Data Release \citep[DR3,][]{GaiaCollaboration2016AA...595A...1G,GaiaCollaboration2023AA...674A...1G}{}. 
The HST observations do not cover the entire field, while the \textit{Gaia} data offer a more homogeneous coverage of the field around Westerlund~1, but only with relatively bright sources. This means that some frames of the short wavelength (SW) channel, which cover a smaller field of view compared with those in the long wavelength (LW) channel, only have \textit{Gaia} sources available, many of which are saturated in our NIRCam data. As the alignment of the LW images to the reference catalog was successful, we used a narrow-band filter, because stars saturate at brighter magnitudes than when wide filters are used. We thus first aligned the F323N filter, extracted a source catalog, and used that catalog as a reference to align the rest of the filters. The details of the final adopted procedure are as follow: 
\begin{itemize}
    \item We ran \textit{tweakreg} on each F323N Stage 2 \textit{cal} exposure individually, extracting the sources from the \textit{cal} images using the DAOFIND algorithm \citep{Stetson1987PASP...99..191S} as implemented in \texttt{astropy} and using the HST+Gaia catalog as reference.
    \item After each frame was aligned, we added the newly found sources to the reference catalog in order to improve its completeness and homogeneity.
    \item The final mosaic was then obtained using the aligned frames with the standard Stage 3 pipeline with \textit{tweakreg} deactivated.
    \item We obtained a catalog of the entire F323N mosaic using DOLPHOT (as described below).
    \item We then aligned the rest of the filters using the F323N source catalog as reference. As for the F323N filter, the frames were individually aligned using \textit{tweakreg}, with the alignment option disabled during mosaic production.
\end{itemize}

A number of artifacts\footnote{\url{https://jwst-docs.stsci.edu/known-issues-with-jwst-data/nircam-known-issues/nircam-scattered-light-artifacts\#NIRCamScatteredLightArtifacts-tadpoles\_shells}} were masked in some LW exposures before the combination, namely: strong, so-called ``dragon breath'' artifacts present in two subpixel dithers of module A images of all the LW filters, and a number of ``ghosts'' present in the two subpixel dithers of two pointings in both module A and B and in all narrow LW filters (F323N, F405N and F466N). In the F277W observations, there were also a number of ghosts that were masked. Masking was only performed if there were available data at the position of the artifact in another pointing. \\

Fig.~\ref{Fig:nircam_mosaic} shows the resulting NIRCam RGB mosaic produced by combining the images with the filters F277W (blue), F323N (green), and F444W (red). We note the intricate superposition of diffraction spikes generated by the bright stars in the center, which limits the completeness toward faint sources in the cluster core. NIRCam mosaics also show some of the nebulosity surrounding Westerlund~1, which mainly emits at longer wavelengths; this is discussed in more detail below.  

   \begin{figure*}
   \centering
   \includegraphics[width=1\textwidth]
   {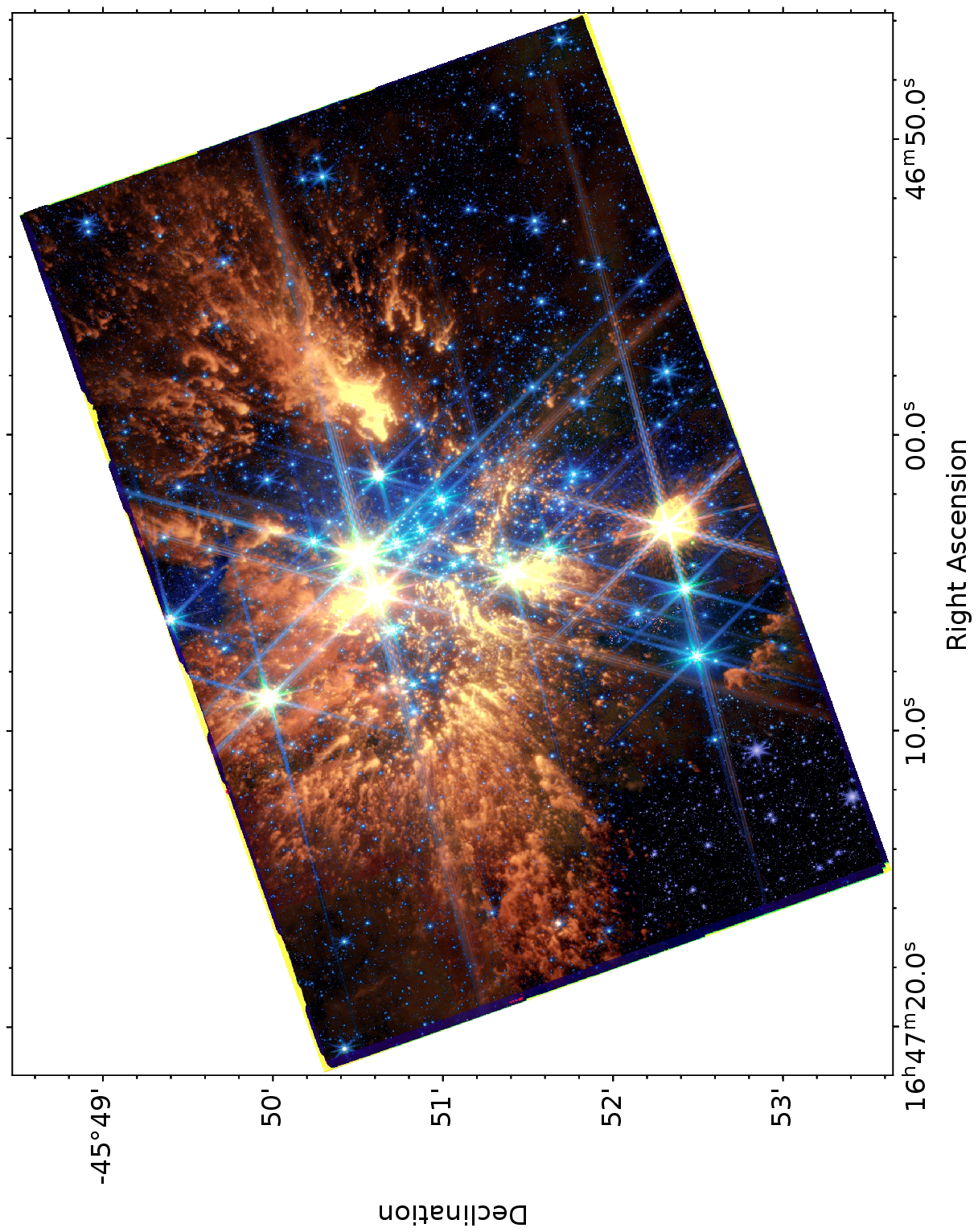}
      \caption{Combined NIRCam/MIRI RGB image of Westerlund 1 (red: F1130W, green: F770W, blue: F444W). At 4230$\,$pc, the angular size of this image spans an area of 4.3$\times$4.9$\,$pc.}
         \label{Fig:nircam_miri_mosaic}
   \end{figure*}

\subsubsection{MIRI}
\label{sec:mirired}

We observed the Westerlund~1 field and CF with MIRI in three filters, as described in Table~\ref{tab:MIRIobssetup}. 
In the Westerlund~1 field, we observed nine different pointings for each setup. We opted for this setup to ensure excellent sensitivity to extended structures in a small number of pointings spanning the extended field of Westerlund~1, whilst aiming to minimize the inevitable saturation from bright stellar sources. 
The \texttt{uncal.fits} files produced by the each pointing were analyzed as described in Sect. \ref{sec:datared}, with Guide Star Catalog version 2431.


The reduction of MIRI data and the production of the mosaic were performed mostly adopting the default settings of the MIRI pipeline, with a few exceptions. For instance, at stage~3, we adopted \texttt{abs\_minobj}=30 and \texttt{abs\_minobj}=40 in the \textit{tweakreg} step, with these parameters indicating the minimum acceptable number of objects that can be used for alignment. 

To confirm the performance of our calibration and imaging pipeline, we first ran this on the MIRI CF (see Table~\ref{tab:MIRIobssetup} and Appendix \ref{appendix:cf}). 
The RGB MIRI map obtained using this procedure, together with the images of the CF obtained with the NIRCam filters F115W and F277W, are presented in Appendix~\ref{appendix:cf}. We thus ran the same pipeline on the MIRI science fields (see Table~\ref{tab:MIRIobssetup}) producing the RGB maps shown in Fig. \ref{Fig:nircam_miri_mosaic}, for example.

\subsection{NIRCam and MIRI photometry}
\label{sec:photometry}

PSF photometry on the NIRCAM and MIRI images was  performed using DOLPHOT \citep{Dolphin2000, Dolphin2016}. DOLPHOT performs PSF photometry on the Stage 2 images (\texttt{cal.fits}) using the stage~3 mosaic as a reference for source detection and image alignment. We use the DOLPHOT version updated on 4 February 2024, which includes the latest PSF libraries, Sirius-Vega zero points, and the \texttt{-etctime} pre-processing flag, which adjusts the exposure times in the header to be in agreement with the exposure time calculator (ETC). The DOLPHOT PSF libraries are calculated using WebbPSF\footnote{\url{https://webbpsf.readthedocs.io/en/stable/index.html}} \citep{Perrin2012SPIE.8442E..3DP,Perrin2014SPIE.9143E..3XP} version 1.2.1. See \citet{Weisz2024ApJS..271...47W} for details on how the DOLPHOT PSF library was built. For NIRcam, we used the parameters recommended in \citet{Weisz2024ApJS..271...47W} and the NIRCAM manual\footnote{\url{http://americano.dolphinsim.com/dolphot/dolphotNIRCAM.eps}}, which includes the PSF and sky fitting parameters, as well as the strategy to perform the aperture correction and image alignment. 

DOLPHOT performs PSF photometry on all the sources above the pre-defined minimum signal--to--noise threshold (sigPSF parameter); this value is fixed at 5, as recommended in \citet{Weisz2024ApJS..271...47W}. As expected, a large fraction of the sources in the output catalog are spurious. For instance, DOLPHOT detects a large number of sources in the saturation spikes of the very bright stars. However, DOLPHOT provides several photometry fit metrics (such as sharpness, roundness, $\chi^2$, crowding, object type, and \textit{photometry quality}) that can be used to identify reliable detection of stellar sources. \citet{Warfield2023RNAAS...7...23W} developed a set of criteria to identify real sources using DOLPHOT for the F090W and F150W NIRCam filters based on these metrics. We tested these criteria on our images in various filters and adopted the same criteria except for the crowding metric, which we restrict to values of lower than or equal to 0.3. We find that the vast majority of sources with crowding between 0.3 and 0.5 (recommended value) are located in the spikes and extended PSF region of the saturated stars. After performing these photometry metric cuts, only $\sim$10\% of the detected sources remain.


\section{Overview of the JWST images of Westerlund 1} \label{sec:features}

The JWST images (Figs. \ref{Fig:nircam_mosaic} and \ref{Fig:nircam_miri_mosaic}) display various interesting features that will be the subject of forthcoming papers of the EWOCS survey. In the following sections, we discuss the source distribution in the NIRCam color--magnitude diagrams, and the diffuse nebulosity and shells around massive stars observed with MIRI.  

\subsection{The cluster sequence in the NIRCam diagrams}
\label{sec:nircamphot}

The NIRCam observations were designed to provide powerful diagnostics to determine the properties (such as masses and ages) of the population of Westerlund~1 down to the substellar regime. This was achieved from the analysis of the NIRCam color--color and color--magnitude diagrams, accounting for the incompleteness (Sect.~\ref{sec:photometry}) and taking advantage of the \textit{Chandra}/ACIS-I X-ray catalog of the region published in \citet{Guarcello2024AA...682A..49G}. The \textit{Chandra} observations are important because of the well-known high-level of X-ray emission in pre-main sequence stars \citep{Montmerle1996}, which would allow independent selection of the stars associated with Westerlund 1. \par

With this aim, we first identified the NIRCam sources with X-ray counterparts. The spatial density and photometric depth of the NIRCam catalog, when compared to those of the X-ray catalog, necessitate an accurate procedure to match the sources common on the two, which cannot solely rely on a closest-neighbor-based procedure. We thus employed a maximum-likelihood procedure that compares not only source positions, but also the magnitude of candidate counterparts with the magnitude distribution of the expected real optical-infrared counterparts of the X-ray sources. This matching procedure is similar to the one adopted by \citet{Guarcello2023ApJS..269....9G}, which is derived from that of \citet{Smith2011MNRAS.416..857S}, and results in 2170 matches, with 2074 single matches, from the 3888 X-ray sources from \citet{Guarcello2023ApJS..269....9G} that fall within the NIRCam field. Further details about the cross-match between the \texttt{Chandra}/ACIS-I and the NIRCam catalogs are provided in the Appendix \ref{appendix:match}. \par 

   \begin{figure}
   \centering
   \includegraphics[width=0.5\textwidth]{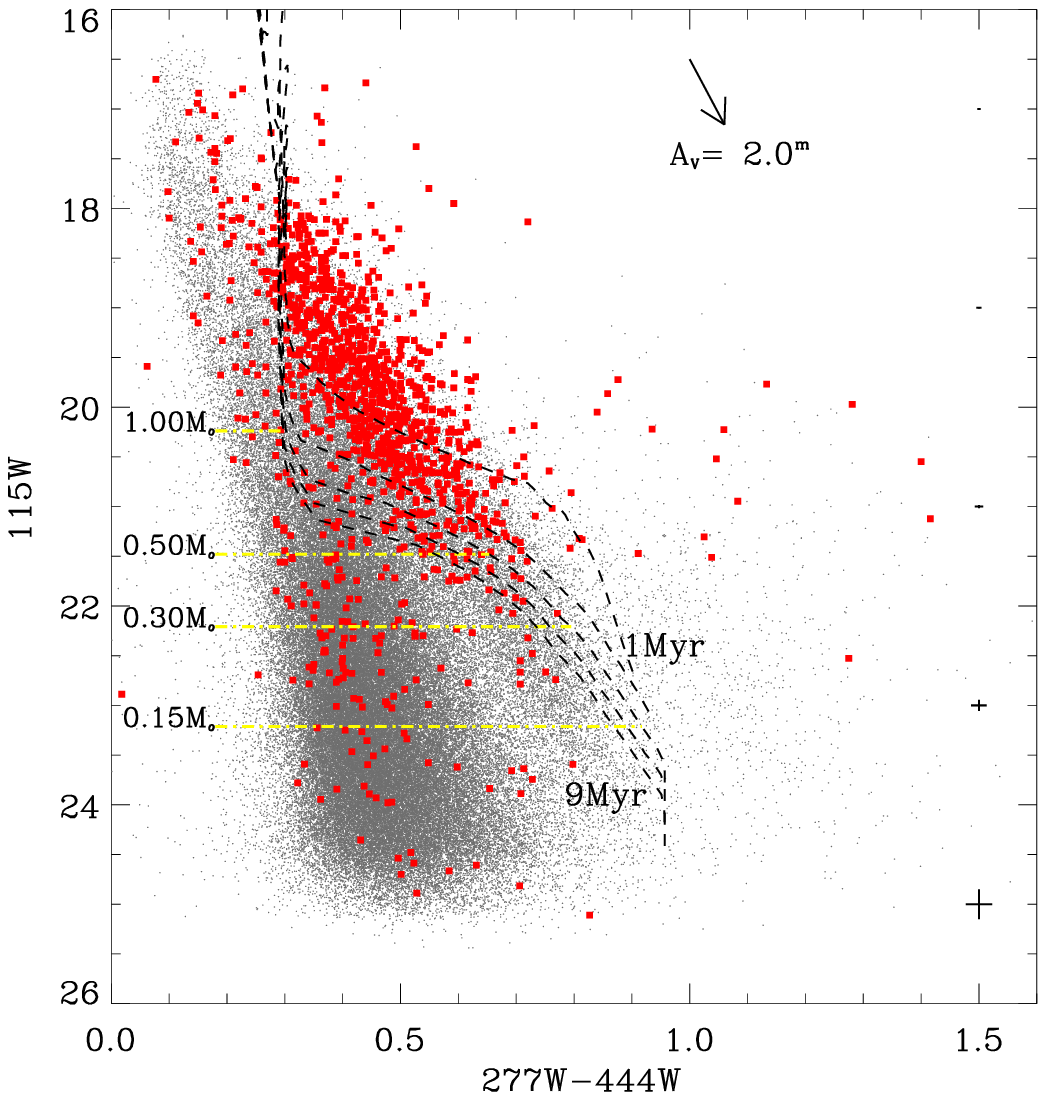}
      \caption{NIRCam color--magnitude diagram of the sources in the Westerlund~1 field (gray dots). The red symbols mark the colors and magnitudes of the NIRCam sources with an X-ray counterpart. The dashed lines are PARSEC isochrones with ages of 1 (top), 3, 5, 7, and 9$\,$Myr (bottom), drawn adopting a distance of 4230$\,$pc and A$\rm_V$=10$\,$mag. The magnitudes corresponding to the masses on the left are taken from the 5$\,$Myr isochrones. The extinction vector is drawn assuming the color excesses provided by the PARSEC website. The crosses on the right show the median error bars in bins of 2 mag in width (e.g., the first, barely visible, point accounts for stars in the 16-18 mag bin).}
         \label{fig:cmd}
   \end{figure}

   \begin{figure}
   \centering
   \includegraphics[width=0.5\textwidth]{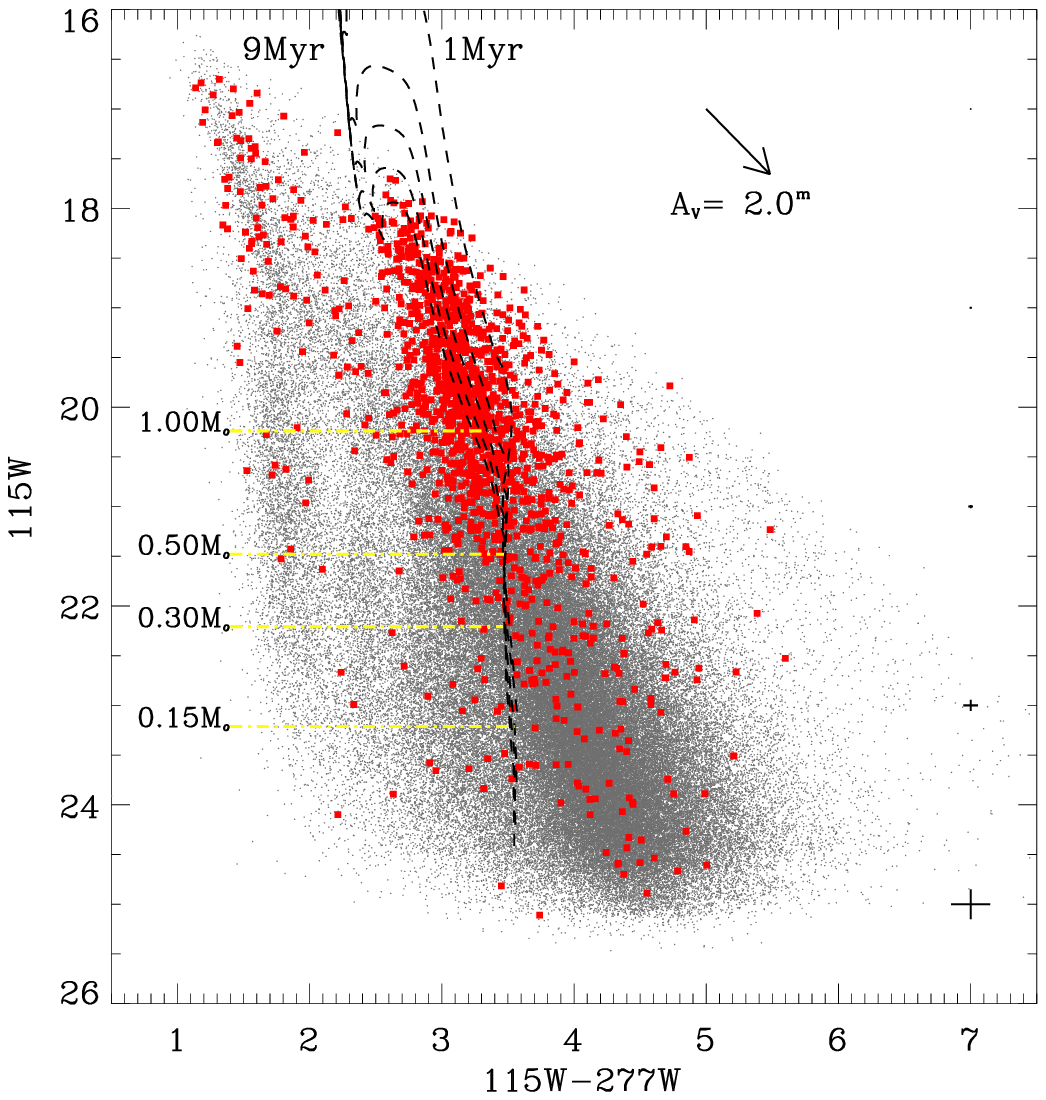}
   \includegraphics[width=0.5\textwidth]{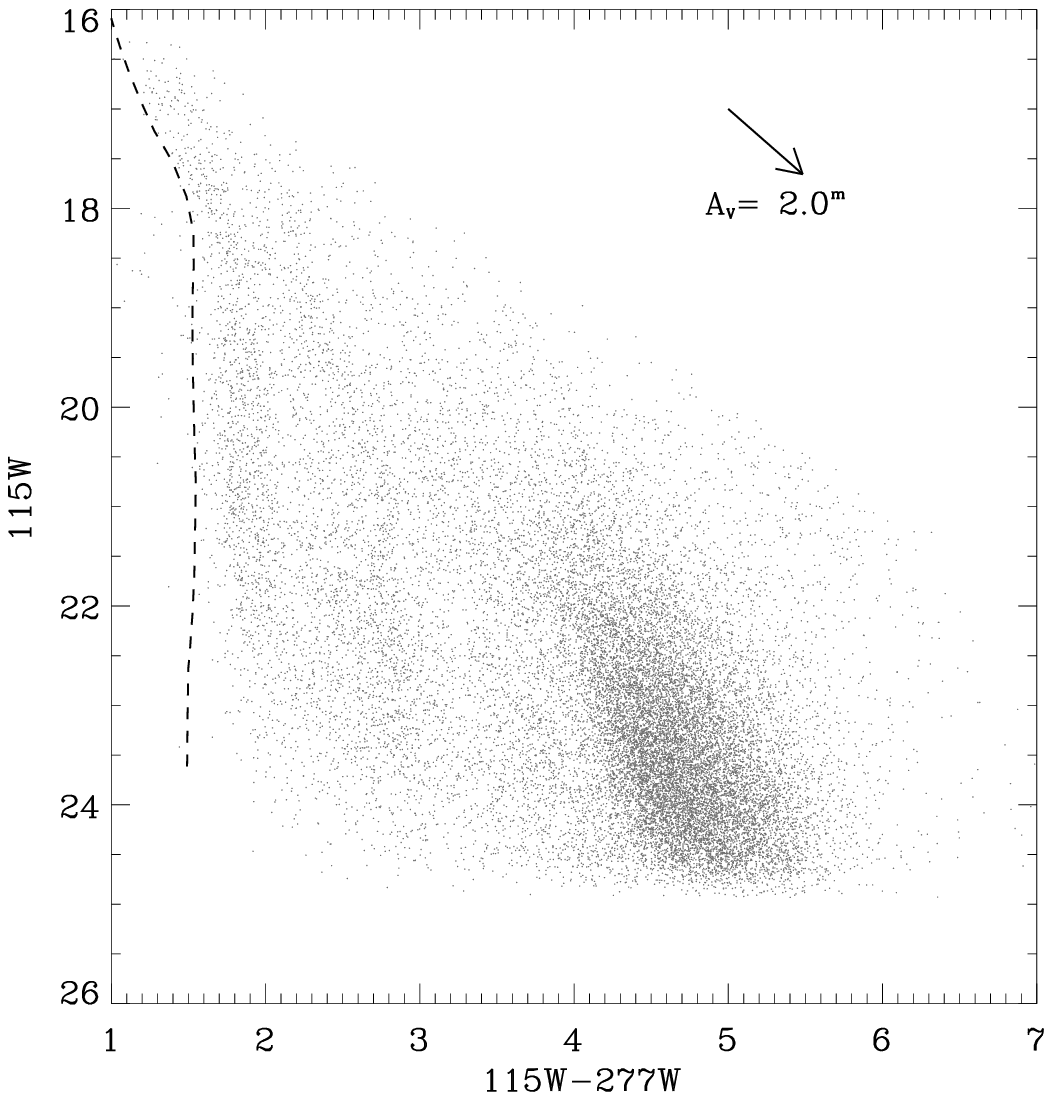}
      \caption{NIRCam color--magnitude diagram of the sources in the Westerlund~1 field (top panel) and in the control field (bottom panel). Symbols as Fig. \ref{fig:cmd}. In the bottom panel we show the 5$\,$Gyrs PARSEC isochrone at a distance of 2$\,$kpc and extinction of 2 mag.}
         \label{fig:cmd2}
   \end{figure}

The NIRCam color-magnitude diagram drawn with the F115W, F277W, and F444W filters from the whole NIRCam field is shown in Fig. \ref{fig:cmd}. The isochrones with ages of between 2 and 10$\,$Myr are obtained from the PARSEC\footnote{\url{http://stev.oapd.inaf.it/cgi-bin/cmd}} database \citep{Bressan2012MNRAS.427..127B} and are plotted adopting a distance of 4230$\,$pc and an extinction of A$\rm_V$=10$\,$mag. \par

In the magnitude range of 18$<$F115W$<$21, the locus of the stars associated with Westerlund~1 members is marked by the NIRCam sources with an X-ray counterpart. The lower limit of this locus approximately corresponds to stars between 0.5$\,$M$_\odot$ and 1$\,$M$_\odot$ at the distance and extinction of the cluster. The lower limit is not well defined because of the incompleteness of the X-ray catalog, and the variability of X-ray emission, as the detection in X-rays for the low-mass cluster members strongly depends on the flaring activity. A few NIRCam+X-ray sources are brighter and bluer than the cluster locus, while a few tens of these sources populate the locus with F115W$>$22 and F277W-F444W$<$0.8. The former are likely sources in the foreground, which is not surprising considering the distance to Westerlund~1, and the fact that part of the population of the young cluster BH197 (at a distance of about 2$\,$kpc) falls within the ACIS and NIRCam fields. The latter group of NIRCam+X-ray sources are mainly located far from the cluster core, and could be real background X-ray emitting sources, observed at large extinction, or spurious coincidences between X-ray and background NIRCam sources. \par

The extension of the cluster locus as marked by the NIRCam+X-ray sources along the extinction vector, and the shape of the isochrones at low masses, seem to indicate that the diagonal population with F115W$>$21 and 0.6$<$F277W-F444W$<$1.0 could contain most of the cluster members down to the substellar regime. \par

A direct comparison between the Westerlund~1 field and the CF stellar populations can be made using the F115W versus F115W-F277W diagram, shown in Fig. \ref{fig:cmd2}. In the Westerlund~1 field, two foreground stellar populations affected by lower extinction than the cluster members can be observed at F115W-F277W$\sim$2 and F115W-F277W$\sim$2.8. The background population is instead extended for F115W-F277W$\geq$4, following the direction of the extinction vector. By comparing the two diagrams, and considering the almost vertical direction of the isochrones, it is evident that most of the stellar population of Westerlund~1, down to very low masses, lies within 3$\leq$F115W-F277W$\leq$4. \par

An artificial star test (see Sect. \ref{sec:photometry})--performed following a similar approach to that of  \citet{Weisz2024ApJS..271...47W}--in order to estimate the completeness of the catalog  will be described in detail in an upcoming paper dedicated to the analysis of the cluster IMF. The 50\% completeness in the F115W filter approximately corresponds to 23.8 mag, although a single value poorly represents the spatial variation of completeness due to crowding and the bright saturated stars. Comparing these values to the 5 Myr \citet{Baraffe2015AA...577A..42B} isochrone (which includes lower masses than the PARSEC isochrones) shifted to a distance of 4230$\,$pc and with extinction of Av=10 mag, we find that the estimated 50\% completeness corresponds to a mass of 0.06 $M_\odot$.

\subsection{The nebulosity in and around Westerlund~1}
\label{sec:nebulosity}

   \begin{figure*}
   \centering
   \includegraphics[width=0.7\textwidth]{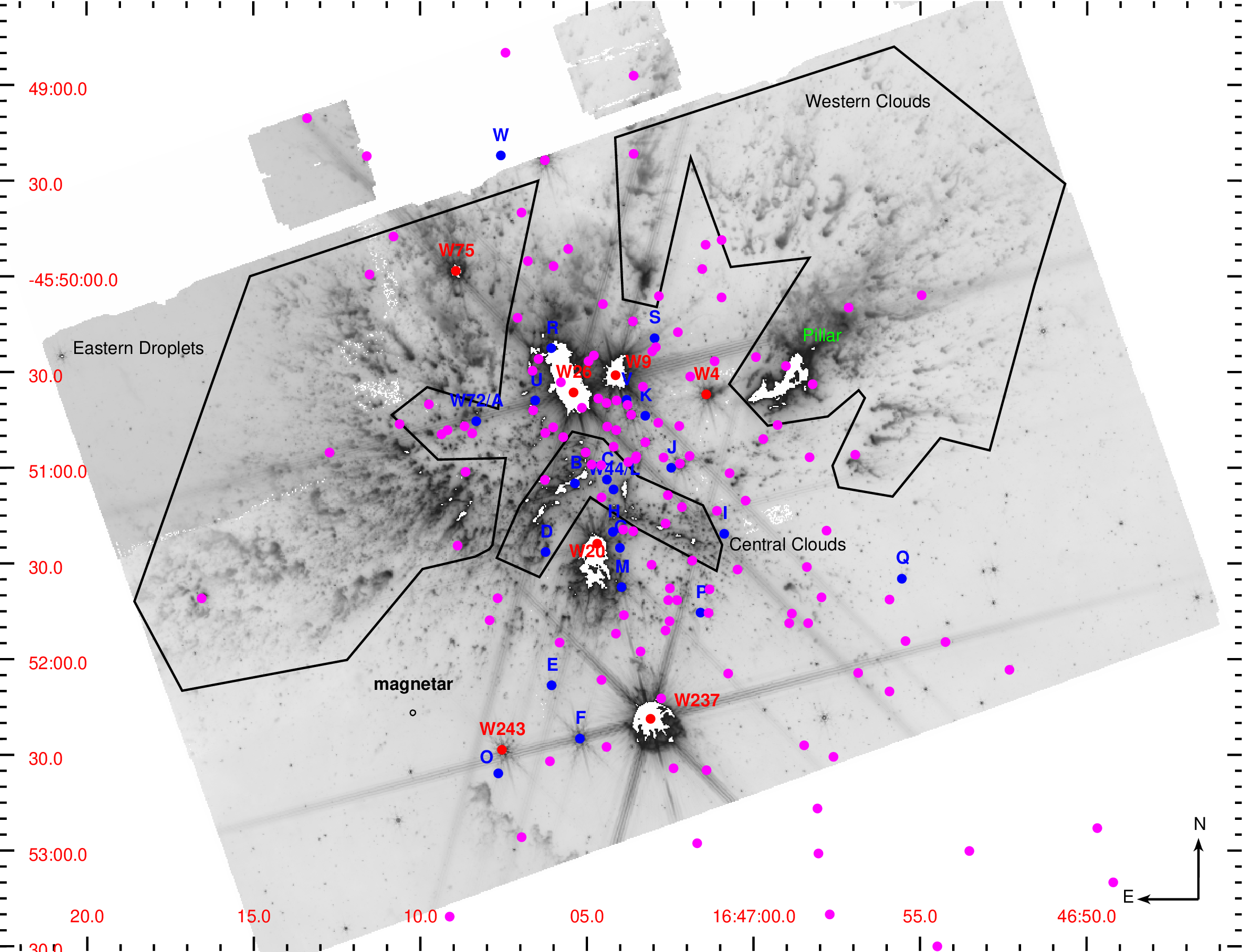}
      \caption{MIRI image of Westerlund~1 taken with the JWST/MIRI F1000W filter. The black polygons delineate the nebulosity regions discussed in the text. Purple circles indicate the positions of the massive stars listed by \citet{Clark2020AA...635A.187C}, with WR star positions highlighted in blue and the evolved stars in red. The position of the magnetar CXO J164710.2-455216 \citep[][Borghese et al., in prep.]{Muno2006ApJ.636L.41M} is also indicated.}
         \label{Fig_nebula}
   \end{figure*}
   
One of the most striking features in the MIRI images of Westerlund~1 is the bright and complex nebulosity visible around and within the cluster core. Such dense nebulosity around Westerlund~1 was not expected, considering that optical and infrared images of star clusters with similar ages, or even younger than Westerlund~1, and with less massive stellar content show how they are capable of clearing large cavities, delimited by an expanding photodissociation front \citep[e.g.,][]{Geen2015MNRAS.448.3248G}. \par 

Fig. \ref{Fig_nebula} displays the MIRI image of the cluster captured with the F1000W filter. The black polygons delineate the nebulosity around the cluster core and separate it into three distinct regions:
\begin{itemize}
    \item The ``Eastern Droplets'' are characterized by droplet-like features with typical projected sizes of a few arcseconds (which corresponds to a projected physical extent of less than a few tenths of parsecs). These droplets are primarily oriented toward the eastern group of massive stars, which contains the WN7b star W72/A, and the X-ray pulsator recently identified in the EWOCS data (Israel et al., in prep.). Additionally, some droplets in the southeast point toward the main core of massive stars, while some in the northwest point toward the space between these two groups of massive stars.
    \item The ``Western Clouds'' exhibit fewer, larger, and more sparsely distributed droplet-like features, along with a large pillar of approximately 25 arcsec in length (roughly corresponding to a projected size of 1 parsec) directed toward the central cluster of massive stars.
    \item The ``Central Clouds'' display shapes that are indicative of intense radiative and mechanical feedback from the massive stars positioned both in the northern and southern directions.
\end{itemize}

In Fig. \ref{Fig_nebula}, the positions of the most massive stars in the cluster are also indicated based on the list published by \citet{Clark2020AA...635A.187C}. Of particular note are the most evolved stars, such as the 24 WRs, the RSGs, the YHGs, and the LBV star W243, which exert the most intense feedback on the surrounding cloud due to their rapid winds and intense UV radiation, and the most evolved supergiants, which are characterized by dense and structured mass loss. Additionally, the position of the magnetar CXO J164710.2-455216 \citep{Clark2014AA...561A..15C,Kavanagh2010int..workE..91K} is indicated. The WR star WR-S in the northwest part of the field has been indicated as the potential companion of its progenitor, suggesting that the supernova explosion occurred somewhere between the two sources. \par

\begin{figure*}
   \centering
    \includegraphics[width=0.49\textwidth]{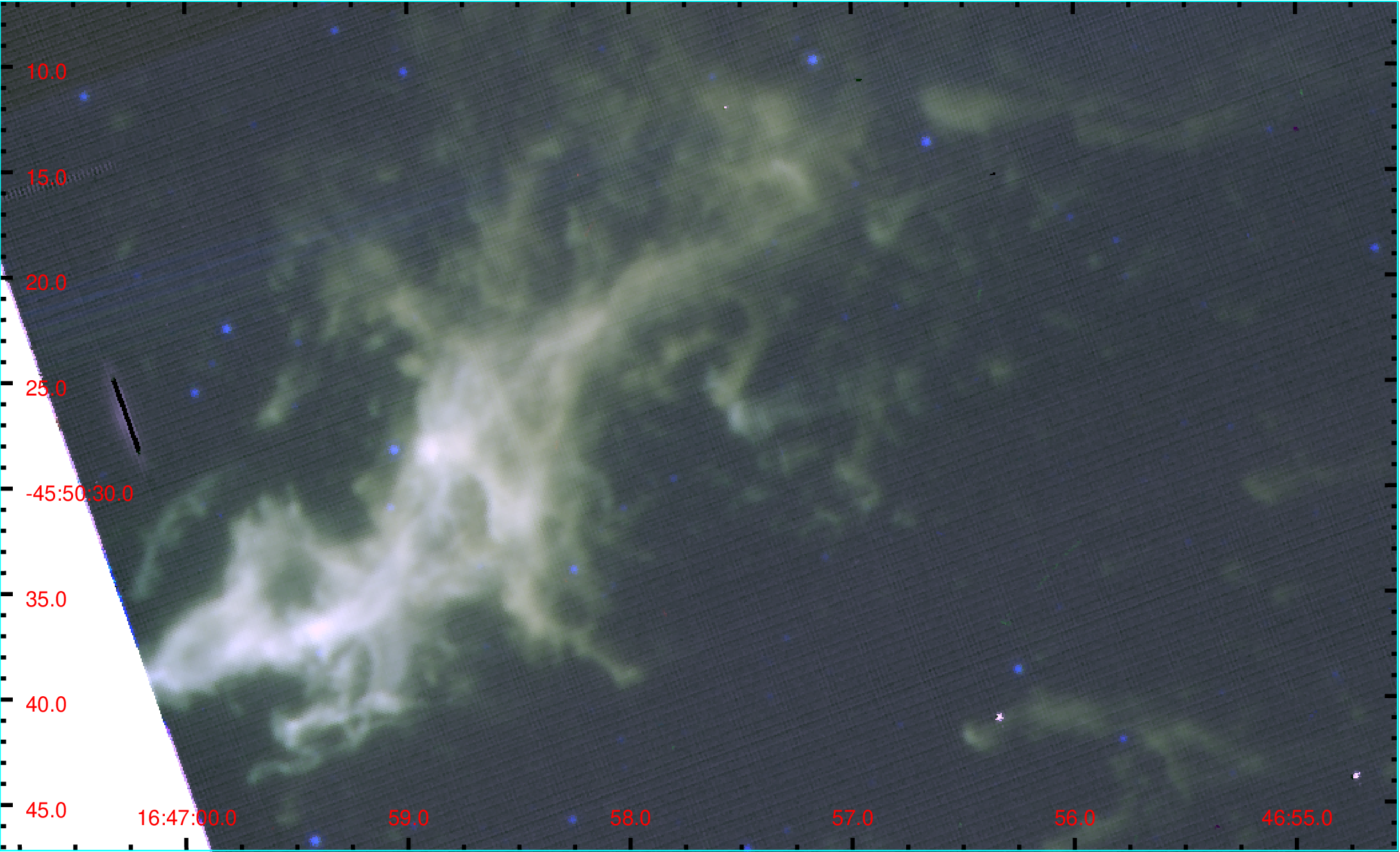}
   \includegraphics[width=0.49\textwidth]{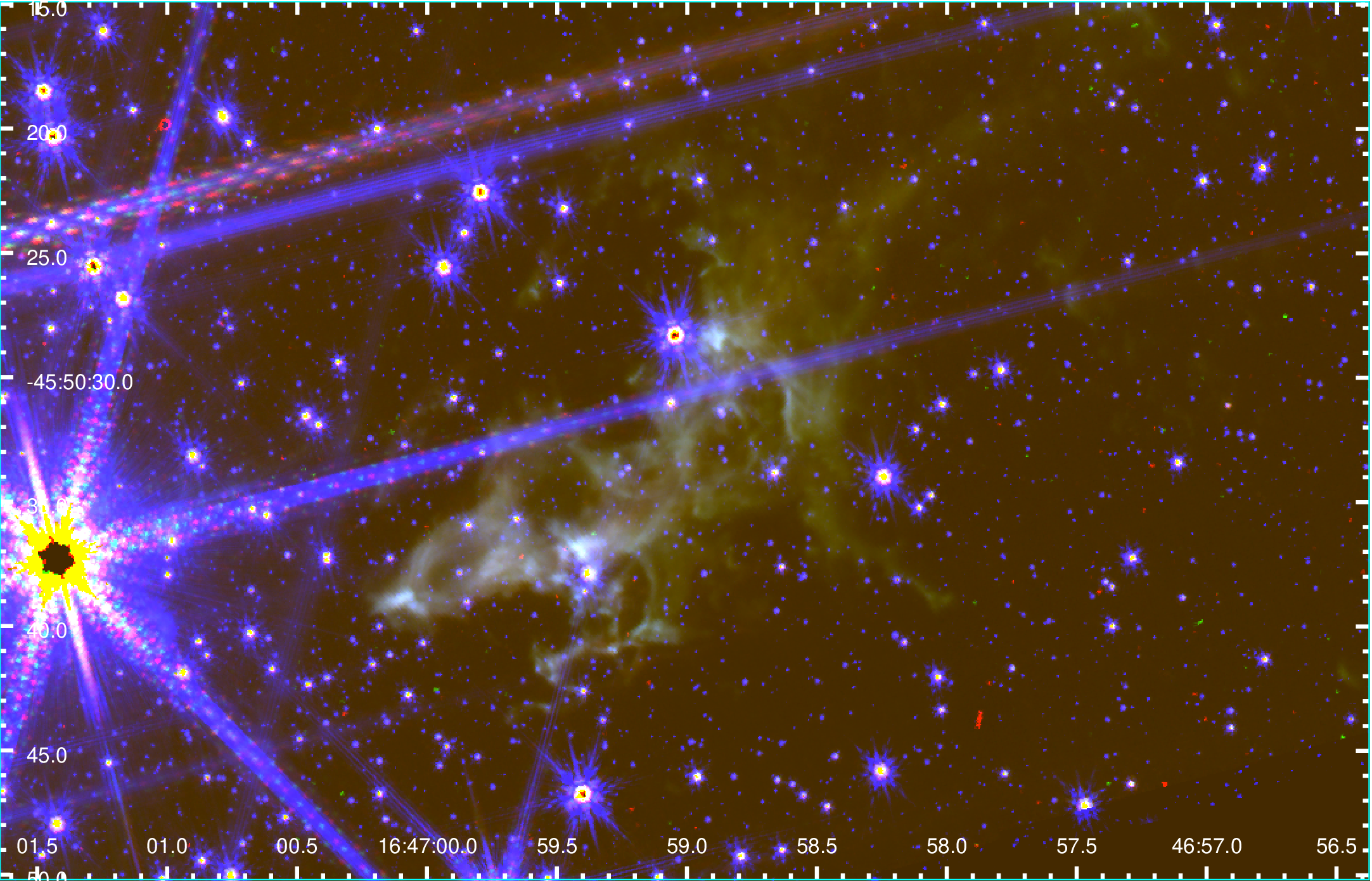}
      \caption{NIRCam RGB images of the pillar in the western cloud. Left panel: Red F466N, green F405N, blue F200W. Right panel: Red F1130W, green F1000W, and blue F770W, created using stage 1 data in order to avoid saturation.}
         \label{Fig_nebulaNIrcam}
   \end{figure*}

Fig. \ref{Fig_nebulaNIrcam} presents two NIRCam RGB images of the elongated trunk in the western cloud. The image in the left panel was composed fromg data captured with the F466N (red), the F405N (green), and the F200W (blue) filters, and shows the contribution of $^{12}$CO (F466N) and Br$\alpha$ (F405N) line emission relative to the continuum emission from warm dust. The image in the right panel displays the pillar from stage 1 data--which are less impacted by saturation--using the filters F770W, F1000W, and F1130W. The absence of saturation in these images along the pillar enables us to the highlight some intriguing details of its tangled structure.

An exploration of the nature of the dense nebulosity will be the main focus of dedicated papers within the framework of the EWOCS project. Such an analysis could in principle unveil important information on the evolution of the cluster and its massive stars, and could also add to the increasing evidence supporting late gas accretion onto protoplanetary discs from the surrounding environment \citep{Gupta2023AA...670L...8G} and its possible contribution to planet formation \citep{Kuffmeier2023EPJP..138..272K} and even the formation of multiple stellar populations in clusters \citep{WinterClarke2023MNRAS.521.1646W}.  The three primary hypotheses regarding the nature of the nebulosity surrounding the cluster are linked to: remnants of the parental clouds of Westerlund~1, winds carrying polluted material from the WR and supergiant stars of the cluster, and ejecta from previous supernova events in the cluster. In the context of the present paper, we only offer qualitative comments on these hypotheses.\par

As previously noted, dense remnants of the parental cloud are not expected at such a close distance from the core of Westerlund~1, considering the intense feedback exerted by its remarkable concentration of massive stars. Moreover, the various observations in the infrared bands have revealed no evidence of ongoing star formation within the cluster or past extended star formation activity in the surrounding area. However, the pillar seems to differ from the other nebular structures in the field in terms of size and structure, which may indicate some connection with the parental cloud of the cluster.\par 

The droplet features observed in most of the cloud surrounding the cluster resemble the ballistic knots observed in the M1-67 cloud surrounding the WR star WR124 \citep{Zavala2022MNRAS.513.3317Z}, recently captured by JWST\footnote{https://webbtelescope.org/contents/media/images/2023/\par 111/01GTWBK1FNPPJSWZ0DK1JB76SN}. The eastern clouds are oriented toward the WN stars W72/A and WRB, which are located at projected distances of about 0.4 and 0.6$\,$pc from the inner front of the droplets, respectively. In this scenario, the eastern cloud could be formed by the interaction of the winds emitted by these stars during their supergiant phase \citep[which is assumed to have lasted about 0.1 Myrs, during which mass was lost at a rate of approximately $\sim$10$^{-4}$ M$_\odot$/yr, with a terminal velocity of about 50 km/s, e.g.][]{PortegiesZwart2010ARAA..48..431P}, with the present-day cluster wind, or by the intense radiation pressure produced by such a rich and dense ensemble of massive stars. The present-day wind is mainly driven by the WR stars, which are characterized by lower mass-loss rates (on the order of $\sim$10$^{-5}$ M$_\odot$/yr) but much faster winds (v$_\infty\sim$2000 km/s). Evidence for a cluster wind has been provided in X-rays by \citet{Muno2006ApJ.636L.41M}, and will be confirmed by future publications of the EWOCS series.  \par

The droplet-like features in the eastern clouds could therefore be explained as the interaction between the winds emitted by the massive stars during their evolution. This hypothesis could be supported by several recent simulations, such as those of \citet{Badmaev2022MNRAS.517.2818B}, which are specifically tailored to the massive population of Westerlund~1, and those of \citet{RodriguezGonzalez2007MNRAS.380.1198R}. Additionally, the formation of droplet-like features occurs naturally when considering the expected clumpiness of the parental cloud impacted by the winds \citep{RogersPittard2013MNRAS.431.1337R}, with the most energetic winds propagating through channels between high-density clumps in the parental cloud. However, these simulations predict very high temperatures for the clumps produced by the colliding winds. Furthermore, as shown by \citet{MartinezGonzalez2015PhDT.......442M}, in a massive environment like Westerlund~1, intracluster dust is rapidly depleted by phenomena such as thermal sputtering. Therefore, a regular source of small dust is necessary to sustain intense infrared emission from the nebulosity, such as that emitted by massive evolved stars and supernova events, which are known to have occurred in Westerlund~1.\par

\subsection{Dusty shells around massive evolved stars}
\label{sec:shells}


\begin{figure*}[]
    \centering
    \includegraphics[width=0.48\textwidth]{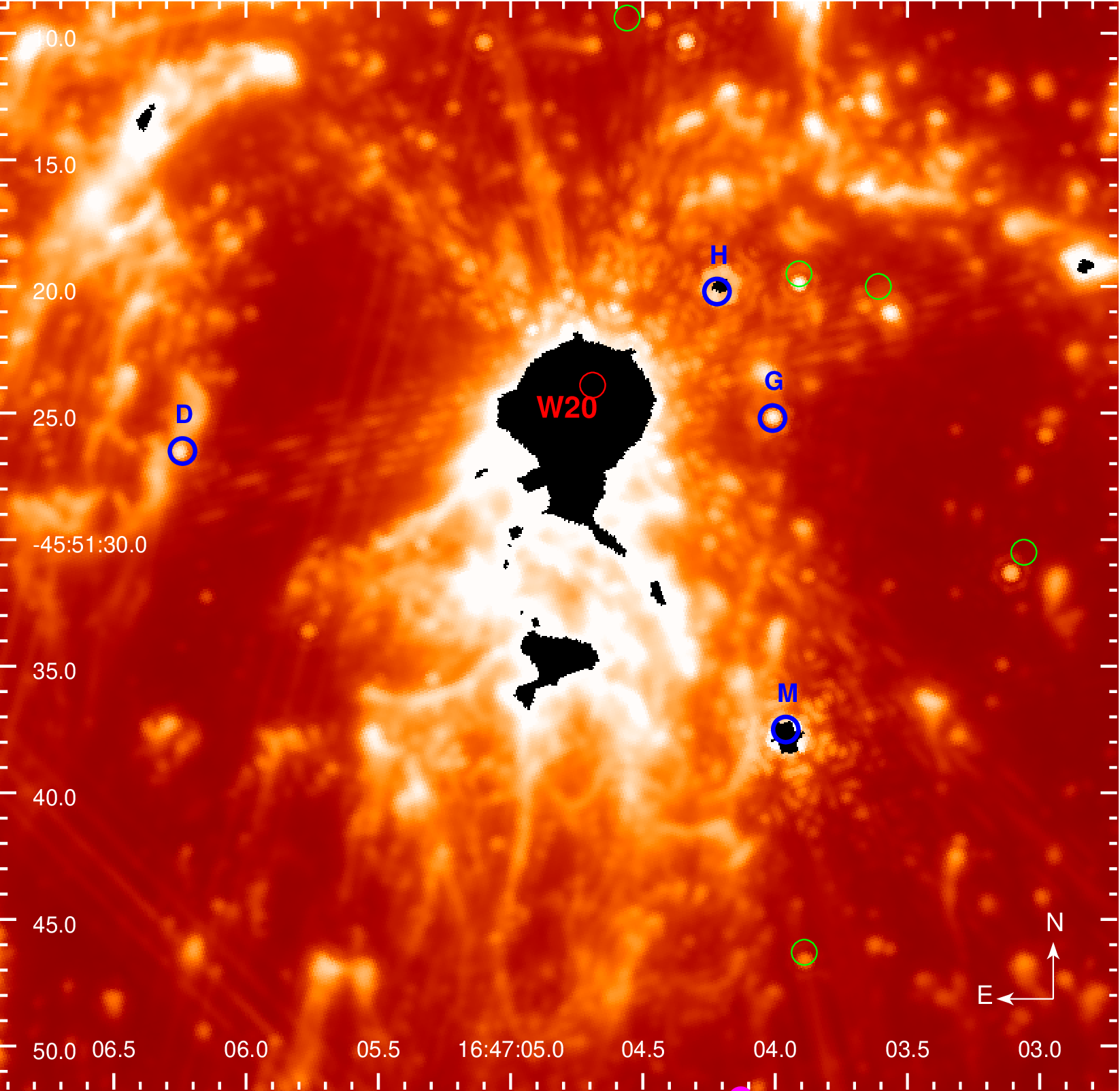}
    \includegraphics[width=0.48\textwidth]{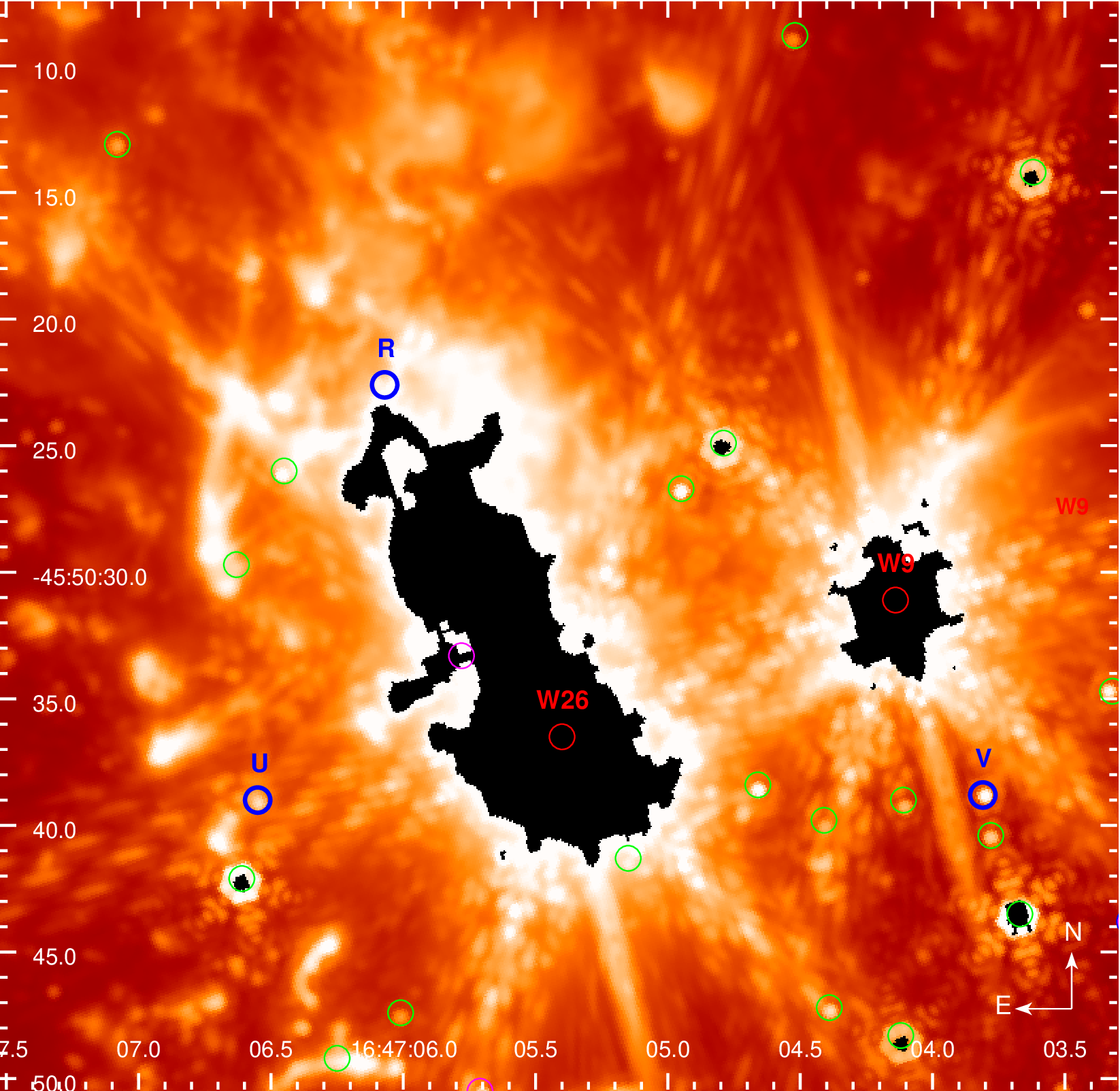}
    \includegraphics[width=0.48\textwidth]{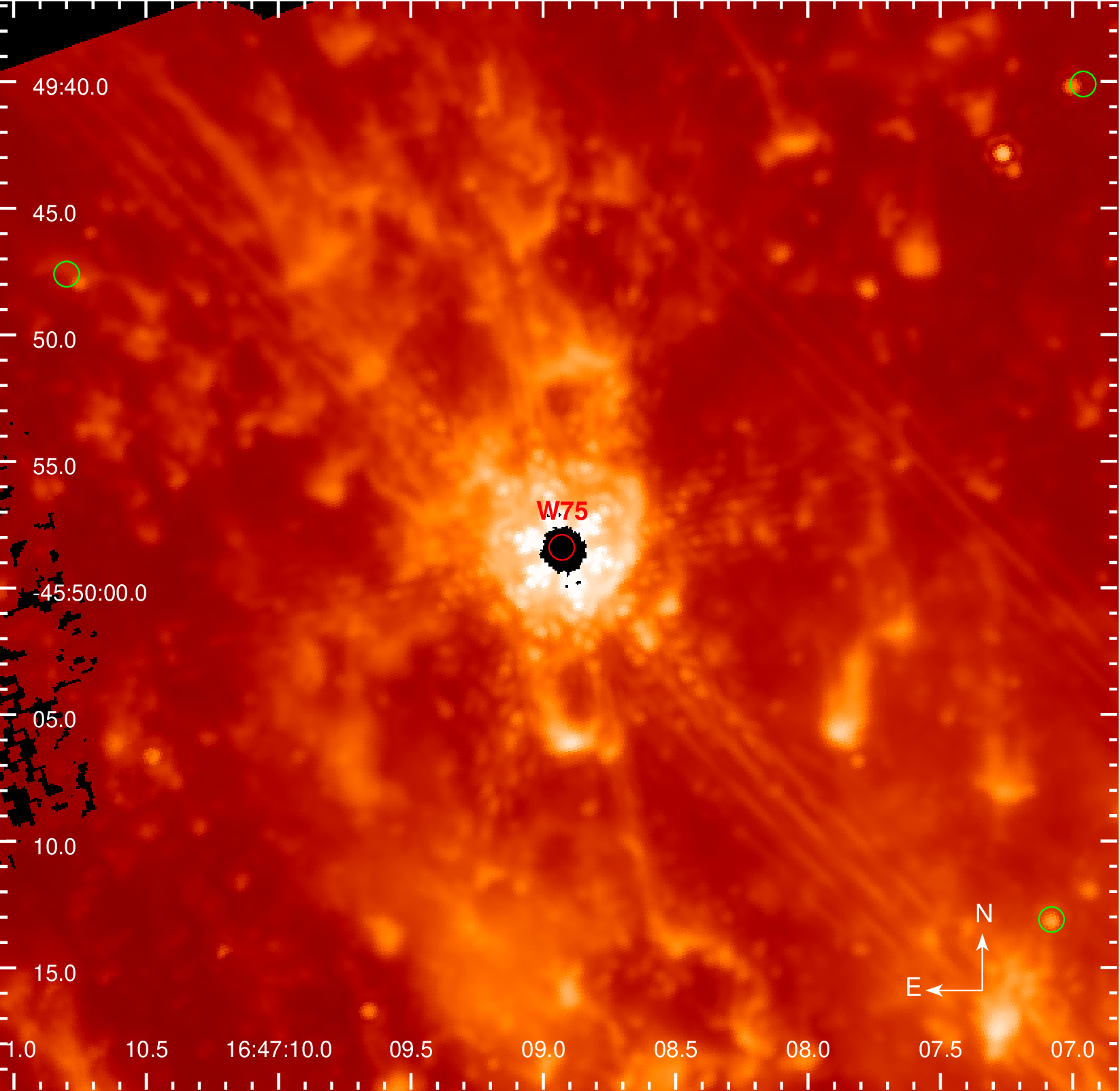}
    \includegraphics[width=0.48\textwidth]{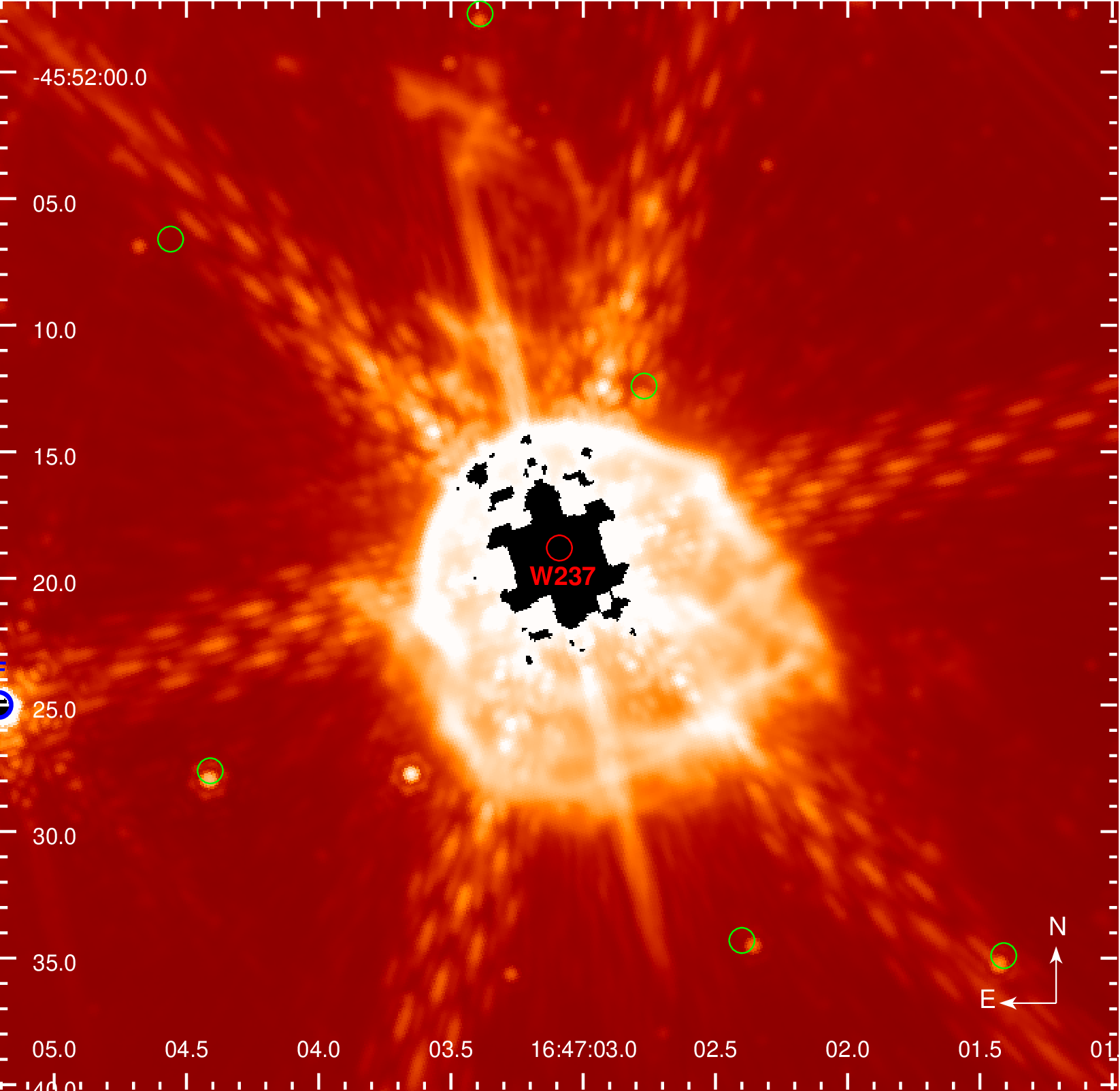}    
    \caption{JWST MIRI images in the F1130W filter of the resolved stellar outflows in the four M-type supergiants of Westerlund 1. In the same panel as W26, the asymmetric outflow from W9 is also visible on the right-hand side of the image. Each panel has a size of 40$^{\prime\prime}\times$40$^{\prime\prime}$.} 
    \label{fig:Mstarsshells}
\end{figure*}

The MIRI images of Westerlund~1 also show detailed shells or outflows surrounding some of the most evolved massive stars in Westerlund 1, such as the four M-type supergiants of the cluster (see Fig. \ref{fig:Mstarsshells}). In some cases, the new observations confirm features observed in radio and millimeter bands by \citet{Andrews2019AA...632A..38A} and \citet{Fenech2018AA...617A.137F}, and offer a high-spatial resolution view that is unmatched by existing observations at other bands. This includes, for instance, the filamentary morphology of the nebulosity around the interacting sgB[e] binary star W9 \citep[the brightest radio source in Westerlund 1;][]{Dougherty2010AA...511A..58D} and the mixed morphology (bow shock plus elongated tail) in the cool supergiants W20 and W237. In other cases, MIRI has unveiled new features not observed previously, such as the bi-lobal ansae towards W75. \\

The MIRI observations also confirm the differences between the outflows of the RSG and those of the YHG stars, as noted by \citet{Andrews2019AA...632A..38A}, strongly supporting the evolutionary differences between these stars in Westerlund 1. The RSGs typically show extended shells with clear evidence of feedback from the surrounding environment. The one exception to this is W75, which shows a very narrow shell close to the star, and bipolar-like lobes whose nature must be addressed with specific follow-up observations. The star W9 on the other hand does not have defined shells, and its outflow, which may come from mass-loss episode during Roche-lobe overflow events, shows a more filamentary structure. \par

   \begin{figure*}
   \sidecaption
   \centering
   \includegraphics[width=0.7\textwidth]{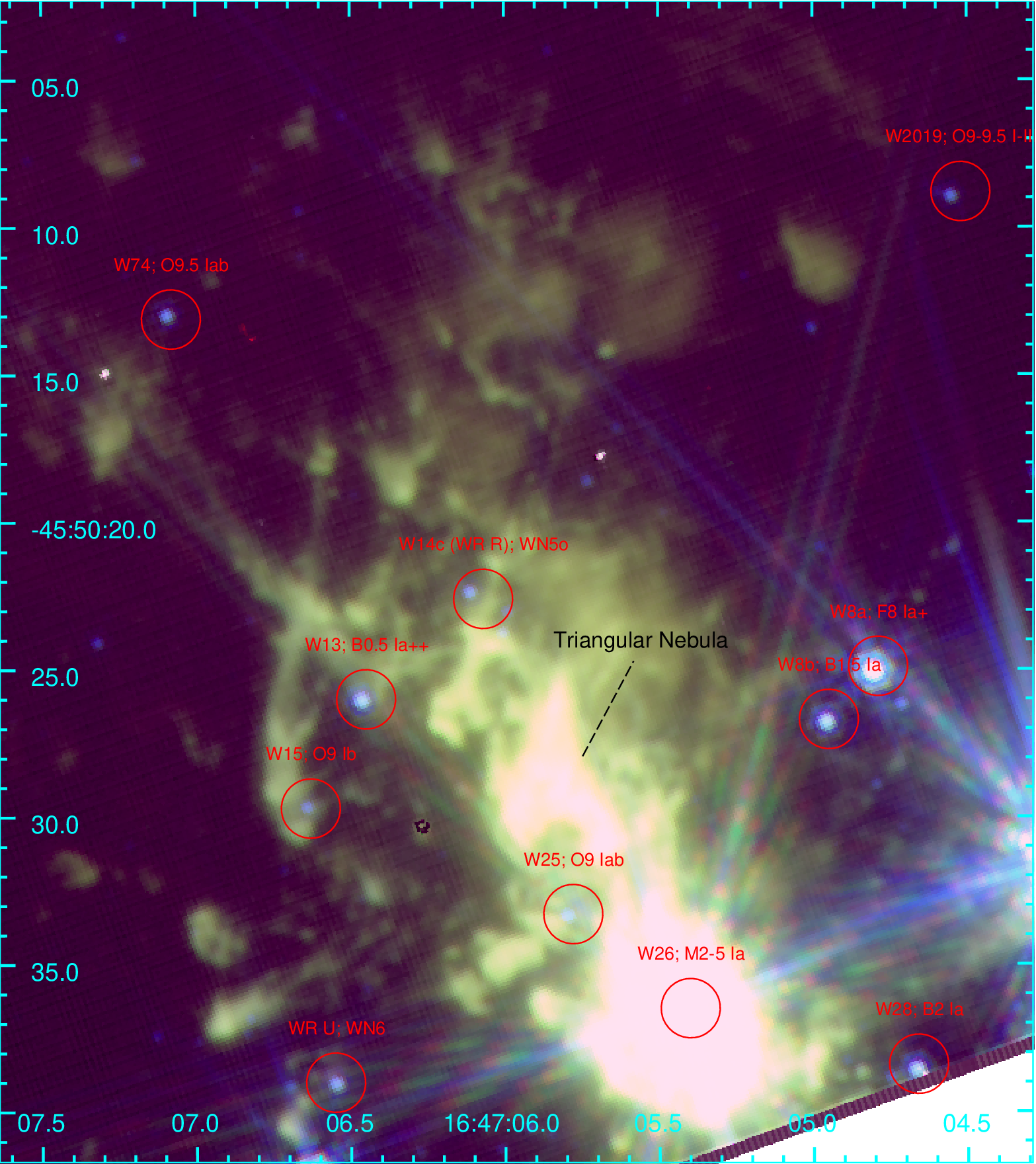}
      \caption{MIRI RGB stage 1 image of the nebulosity associated with the M-type supergiant W26. The positions of the surrounding massive stars, along with the ``triangular nebula'', are marked. }
         \label{Fig:W26}
   \end{figure*}

Fig. \ref{Fig:W26} shows the MIRI RGB image of the nebulosity surrounding the red supergiant W26. We produced this image from stage 1 data, and used it to mitigate the impact of saturation. The nebulosity has a long tail elongated in the opposite direction with respect to the cluster center. These features have previously been described using radio images by \citet{Dougherty2010AA...511A..58D} and \citet{Andrews2019AA...632A..38A}, ALMA 3-mm continuum observations \citep{Fenech2018AA...617A.137F}, optical spectroscopy \citep{Mackey2015AA...582A..24M}, and H$\alpha$ emission from VPHAS observations \citep{Wright2014MNRAS.437L...1W}. Additionally, the association of the outflow of W26 with mid-IR emission was suggested by \citet{Clark1998MNRAS.299L..43C}. The tail is approximately 36$^{\prime\prime}$ long (0.23$\,$pc at a distance of 4230$\,$pc). \par

The nebulosity associated with W26 exhibits two peculiar features. One is the ``triangular nebula'', located approximately 10$^{\prime\prime}$ away from W26 in the northeast direction, which appears to be the largest fragment of the tail. \citet{Wright2014MNRAS.437L...1W} noted that, although the triangular nebula shows a filamentary structure pointing toward the blue supergiant star W25 \citep{Negueruela2010}, it is more likely associated with the outflow of W26 due to the nebular strips connecting the two features. The new MIRI data clearly support this hypothesis. In the MIRI images, the triangular nebula is well defined and bright, and is clearly separated from the nebulosity around the star and from another fragment located to the northeast. \par

The second important feature of this outflow is the evidence of photoionization through UV radiation emitted by the surrounding massive stars, in particular W25 and the two WR stars W14c/R and WR-U. \citet{Mackey2015AA...582A..24M} suggested that the elongated narrow tail of the outflow of W26 could in fact arise from the interaction of the wind of this star with the surrounding ambient radiation. As indicated by \citet{Mackey2014Natur.512..282M}, external ram pressure and UV ionizing radiation from outside, which impacts onto the dense material ejected by RSGs, could in principle confine the winds and allow the accumulation of dense, polluted material in the potential well of the cluster. The strong interaction between the outflow of W26 and the surrounding environment, which has also been suggested by \citet{Fok2012ApJ...760...65F} in order to justify the weak maser activity observed in this star compared with other known RSGs, can thus provide important information on the external feedback acting on the outflows from supergiants and their role in the evolution of the cluster and its star-formation activity. \par


A more thorough analysis of these massive stellar shells and outflows will be presented in an upcoming EWOCS paper.


\section{Conclusions}
\label{sec:conclusions}
The Extended Westerlund 1 and 2 Open Clusters Survey (EWOCS) is a program designed to study star and planet formation and stellar evolution in starburst regions adopting as test cases the two closest supermassive star clusters to the Sun: Westerlund 1 and Westerlund 2. In this paper, which is the third publication of the EWOCS series, we present the \textit{James Webb} Space Telescope NIRCam and MIRI observations of Westerlund 1 (GO-1905). The observations are designed to maximize the sensitivity toward the stars with disks and the low-mass stars of the cluster, as far as the brown dwarf regime. \par

The NIRCam mosaic covers an area of approximately $6^\prime\times6^\prime$ centered on the cluster, while MIRI covers a smaller area of $3.5^\prime\times5^\prime$. Data reduction was performed adopting the standard JWST pipeline, together with tools designed to solve specific issues such as the 1/f noise in NIRCam images. Source detection and photometry were performed with Dolphot. \par

The color--magnitude diagrams produced with NIRCam filters show how the observations succeed in reaching the brown dwarf regime in the cluster. A preliminary estimate of the average completeness (e.g., ignoring spatial variations) suggests a 50\% completeness at 23.8 mag in the F115W, roughly corresponding to a 0.06$\,$M$_\odot$ source in the 2$\,$Myr isochrone at 4230$\,$pc and A$\rm_V$=10 mag. The high photometric quality of NIRCam observations makes these color--color and color--magnitude diagrams excellent tools for separating the population of Westerlund 1 from stars in the foreground and in the background.\par

The MIRI images of Westerlund 1 are dominated by a diffuse nebulosity surrounding the cluster core. The vast majority of the nebulosity is composed of droplet-like features pointing toward the groups of massive stars (mainly the cluster core dominated by the supergiants W9 and W26 and an apparently separated group of stars dominated by the WR star W72/A), an elongated trunk with a projected size of approximately 1 parsec, pointing toward the cluster center, and a small group of cloud fragments wrapping around the group of massive stars surrounding W20. This diffuse nebulosity, which shows clear evidence of a high level of feedback from the massive stellar winds and radiation, will be the target of a follow-up analysis to determine its composition and dynamics, with the ultimate objective of discerning its nature. \par

A further serendipitous discovery these MIRI images of Westerlund 1 is the presence of extended shells surrounding the M-type supergiants W20, W26, W75, and W237, the yellow hypergiant W4, and the sgB[e] star W9. In W20, W26, and W237, the shells have an elongated structure in the opposite direction with respect the cluster center. W237 also shows a well-defined bow shock facing the cluster core, while in the elongated nebulosity of W26 there are some bright overdensities such as the triangular nebula already observed in H$\alpha$. W4, W75, and W9 do not show large shell-like circumstellar nebulosity, but mainly narrow outflows in specific directions: two opposite lobe-like ansae in W75 and two V-shaped outflows in W9. \par

\begin{acknowledgements}
We appreciate the time and effort donated by our anonymous referee to provide supportive comments on this paper. 
This work is based on observations made with the NASA/ESA/CSA James Webb Space Telescope. The data were obtained from the Mikulski Archive for Space Telescopes at the Space Telescope Science Institute, which is operated by the Association of Universities for Research in Astronomy, Inc., under NASA contract NAS 5-03127 for JWST. These observations are associated with program \#1905.
M. G. G. and V. A. acknowledge the INAF grant 1.05.12.05.03.
J.B.L. acknowledges the Smithsonian Institute for funding via a Submillimeter Array (SMA) Fellowship.
J.B.L was also supported via NASA JWST grant JWST-GO-3523. 
K.M.$^2$ was supported by NASA JWST grant JWST-GO-1905.
K.M.$^3$ acknowledges support from the Fundação para a Ciência e a Tecnologia (FCT) through the CEEC-individual contract 2022.03809.CEECIND and research grants UIDB/04434/2020 and UIDP/04434/2020.     
A.G. acknowledges support from the NSF under grants AAG 2008101 and CAREER 2142300. 
A.J.W. has received funding from the European Union’s Horizon 2020 research and innovation programme under the Marie Skłodowska-Curie grant agreement No 101104656.
A.B. acknowledges Deutsche Forschungsgemeinschaft (DFG, German Research Foundation) under Germany´s Excellence Strategy – EXC 2094 – 390783311.
I.N. is partially supported by the Spanish Government Ministerio de Ciencia, Innovaci\'on y Universidades and Agencia Estatal de Investigación (MCIU/AEI/10.130 39/501 100 011 033/FEDER, UE) under grant PID2021-122397NB-C22, and also supported by MCIU with funding from the European Union NextGenerationEU and Generalitat Valenciana in the call Programa de Planes Complementarios de I+D+i (PRTR 2022), project HIAMAS, reference ASFAE/2022/017. E.S. is supported by the international Gemini Observatory, a program of NSF NOIRLab, which is managed by the Association of Universities for Research in Astronomy (AURA) under a cooperative agreement with the U.S. National Science Foundation, on behalf of the Gemini partnership of Argentina, Brazil, Canada, Chile, the Republic of Korea, and the United States of America.
R.B. acknowledgs support by the INAF Mini-Grant “Physical properties of Accreting young stellar objects: exploration of their light Curves and Emission (PACE)”
Support for program GO-1905 was provided by NASA through a grant from the Space Telescope Science Institute, which is operated by the Association of Universities for Research in Astronomy, Inc., under NASA contract NAS 5-03127.
F.N., acknowledges support by grant PID2022-137779OB-C41 funded by MCIN/AEI/10.13039/501100011033 by "ERDF A way of making Europe"

\end{acknowledgements}

\newpage
\addcontentsline{toc}{section}{\bf Bibliography}
\bibliographystyle{aa}
\bibliography{sample631}



\begin{appendix}


\section{NIRCam and MIRI observations log}

Table \ref{tab:MIRIobssetup} shows the log of the NIRCam and MIRI observations of Westerlund 1

\begin{table*}[h]
    \centering
        \caption{Description of NIRCam and MIRI observations and setup.}

    \begin{tabular}{l|c|c|c|c|c|c|c|c}
    \hline
    \hline
        Obs. date & Field & Filter & Sub-- &  Ints/Groups/ & Dithers & Mosaic & Exposure & Pointing (RA; Dec)  \\
        && & array&Frames [\#/\#/\#] & [\#] & tiles [\#] & time [s] & [ICRS] \\
        \hline
        \multicolumn{9}{c}{NIRCam}  \\
        \hline
        31 Mar 2023& SCI & F405N & Full & 1/2/1 & fullbox & 3 & 525.37 & 16:47:03.998; --45:51:05.03 \\ 
        31 Mar 2023& SCI & F323N & Full & 1/2/1 & fullbox & 3 & 525.37 & 16:47:03.998; --45:51:05.03 \\ 
        31 Mar 2023& SCI & F466N & Full & 1/2/1 & fullbox & 3 & 525.37 & 16:47:03.998; --45:51:05.03 \\ 
        31 Mar 2023& SCI & F115W & Full & 1/2/1 & fullbox & 3 & 525.37 & 16:47:03.998; --45:51:05.03 \\ 
        31 Mar 2023& SCI & F277W & Full & 1/2/1 & fullbox & 3 & 525.37 & 16:47:03.998; --45:51:05.03 \\
        31 Mar 2023& SCI & F444W & Full & 1/2/1 & fullbox & 3 & 525.37 & 16:47:03.998; --45:51:05.03 \\ 
        31 Mar 2023& SCI & F150W & Full & 1/2/1 & fullbox & 3 & 525.37 & 16:47:03.998; --45:51:05.03 \\ 
        31 Mar 2023& SCI & F200W & Full & 1/2/1 & fullbox & 3 & 525.37 & 16:47:03.998; --45:51:05.03 \\ 
        31 Mar 2023& SCI & F187N & Full & 1/2/1 & fullbox & 3 & 525.37 & 16:47:03.998; --45:51:05.03 \\ 
        31 Mar 2023& SCI & F212N & Full & 1/2/1 & fullbox & 3 & 525.37 & 16:47:03.998; --45:51:05.03 \\ 
        31 Mar 2023& SCI & F164N & Full & 1/2/1 & fullbox & 3 & 525.37 & 16:47:03.998; --45:51:05.03 \\ 
    \hline
        14 Sep 2022& CF  & F115W & Full & 1/2/1 & Intramodule & 1 & 171.79 & 16:47:43.190; --46:03:47.00 \\ 
        14 Sep 2022& CF  & F277W & Full & 1/2/1 & Intramodule & 1 & 171.79 & 16:47:43.190; --46:03:47.00 \\  
    \hline    
    \hline
        \multicolumn{9}{c}{MIRI}  \\
        \hline
        04 Apr 2023& SCI & F770W & Full & 4/15/1 & 4 point sets & 9 & 6293.8 & 16:47:03.998; --45:51:05.03 \\ 
        04 Apr 2023& SCI & F1000W & Full & 4/15/1 & 4 point sets & 9 & 6293.8  & 16:47:03.998; --45:51:05.03 \\ 
        04 Apr 2023& SCI & F1130W & Full & 4/15/1 & 4 point sets & 9 & 6293.8  & 16:47:03.998; --45:51:05.03 \\ 
        \hline
        03 Apr 2023& CF & F770W & Full & 4/15/1 & 4 point sets & 1 & 699.3 & 16:47:43.190; --46:03:47.00 \\ 
        03 Apr 2023& CF & F1000W & Full & 4/15/1 & 4 point sets & 1 & 699.3  & 16:47:43.190; --46:03:47.00 \\ 
        03 Apr 2023& CF & F1130W & Full & 4/15/1 & 4 point sets & 1 & 699.3  & 16:47:43.190; --46:03:47.00 \\ 
        \hline 
    \end{tabular}
    \label{tab:MIRIobssetup}
\end{table*}

\newpage
$ $
\newpage

\section{Control field images}

\begin{figure*}
   \centering
   \includegraphics[width=0.49\textwidth]{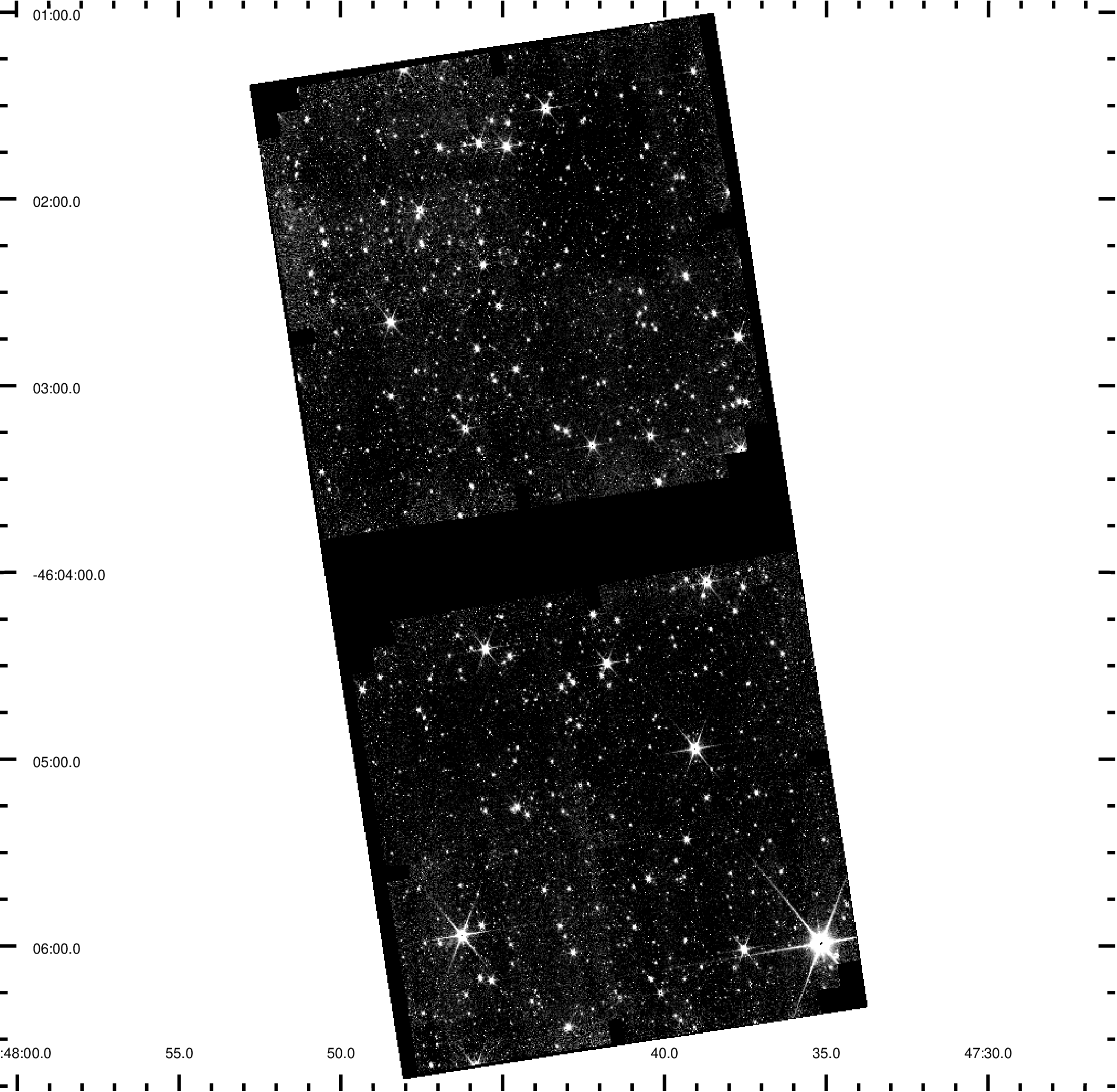}
   \includegraphics[width=0.49\textwidth]{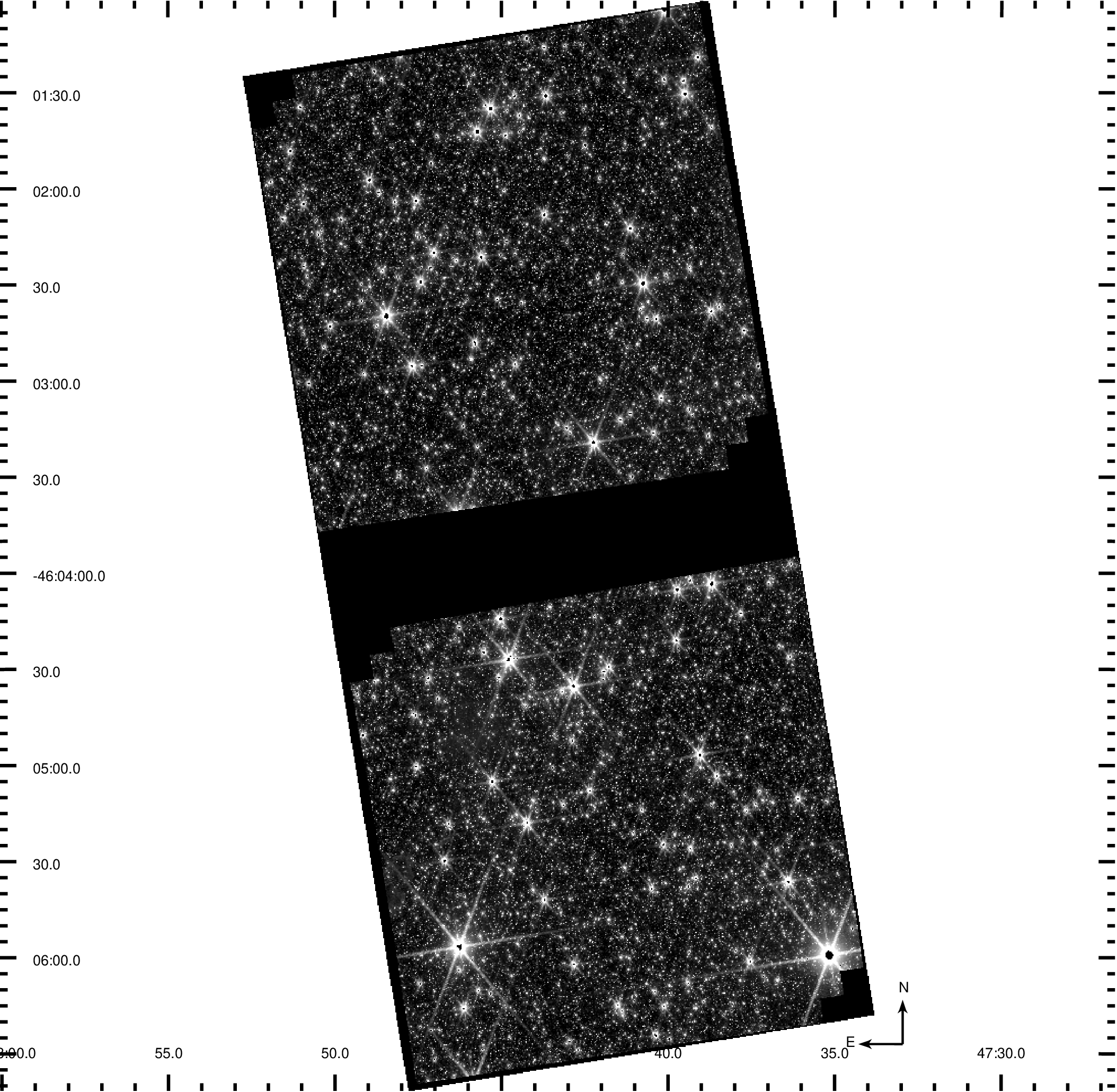}
   \caption{NIRCam images of the control field in the F115W (left) and F277W (right) filters.}
   \label{Fig:cfnircam}
   \end{figure*}

\begin{figure*}[!h]
   \centering
   \includegraphics[width=0.7\textwidth]{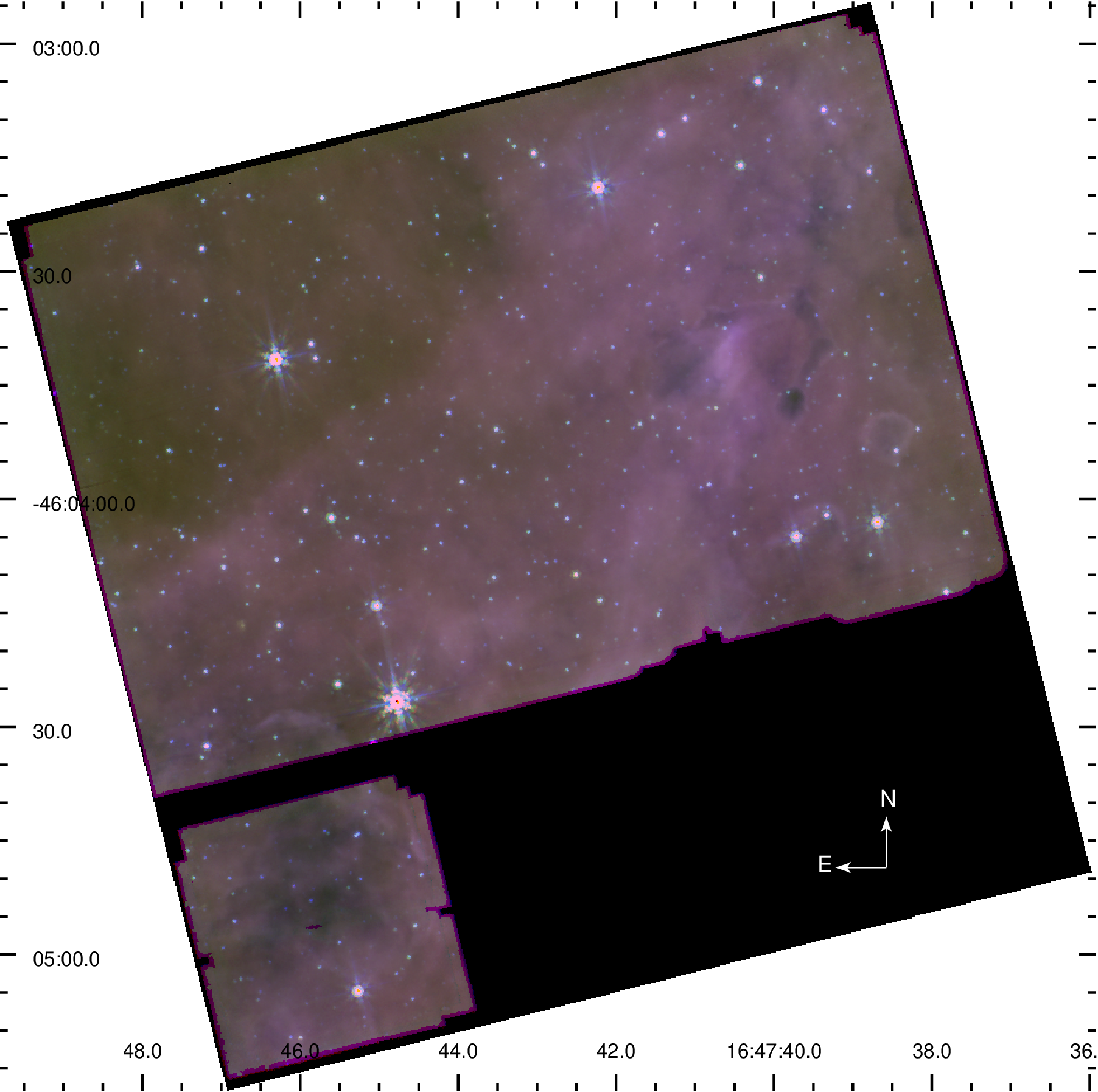}
   \caption{MIRI RGB images of the control field in the F770W (blue), F1000W (green) and F1130W (red) filters.}
   \label{Fig:cfmiri}
   \end{figure*}

In this appendix we show the NIRCam (Fig. \ref{Fig:cfnircam}) and MIRI (Fig. \ref{Fig:cfmiri}) images of the control field. 

\label{appendix:cf}[!h]

\newpage
$ $
\newpage
$ $
\newpage


\section{Matching procedure between the NIRCam and the \texttt{Chandra}/ACIS-I catalog} \label{appendix:match}

In this appendix we describe the matching procedure adopted to combine the NIRCam with the EWOCS \texttt{Chandra}/ACIS-I catalogs. Before merging, the astrometry of the X-ray catalog was anchored the astrometry of the NIRCam catalog by shifting the positions of the X-ray sources by 0.043 arcsec in RA and 0.014 arcsec in dec. \par

For the match, we adopted the maximum-likelyhood (ML) method used by \citet{Guarcello2023ApJS..269....9G}, which has been modified from the method presented in \citet{Smith2011MNRAS.416..857S} to produce the multiwavelength catalog of the \texttt{Chandra} Cygnus OB2 Legacy Survey \citep{Wright2023ApJS..269....7W}. This method relies on the definition of a magnitude distribution of the expected real optical-infrared counterparts of the X-ray sources. This population is obtained from a close-neighbors approach, in which we searched for and merged the pairs of sources in the two catalogs separated by a spatial offset smaller than an individual tolerance (the matching radius $r_{match}$), which was set equal to half the positional error of the X-ray sources (which dominates over the positional error of NIRCam sources). \par

The main step of the ML method is the calculation of the likelihood ratio for each pair of candidate counterparts, equal to: 

\begin{equation}
    LR=\frac{q(m)f(r)}{n(m)}
    \label{eq:LR}
\end{equation}

where $f(r)$ is the radial distribution function of the separations between the X-ray and the NIRCam sources, calculated as:

\begin{equation}
    f(r)=\frac{1}{5\pi\sigma^{2}_{pos}}exp\left( \frac{-r^{2}}{\sigma^{2}_{pos}} \right)
    \label{eq:fr}
\end{equation}

where r is the positional offset between two sources and $\sigma_{pos}$ the positional uncertainties. The quantities $q(m)$ and $n(m)$ are the magnitude probability distributions of the correlated sources (which are the candidate real counterparts) and the observed magnitude probability distribution of all the NIRCam sources in the wide filters, respectively. The distribution $n(m)$ was calculated directly from the whole NIRCam catalog, using the first band in order of increasing wavelength in which the NIRCam source is detected, and $q(m)$ from the close-neighbors match as described above. \par

After the $LR$ ratio is calculated, it is used to estimate the reliability that the given pair of sources are true counterparts. To do this, we first calculated the distribution of $LR$ values from 200000 test X-ray sources uniformly distributed across the field and matched them with the NIRCam sources. The reliability associated with each pair by definition is the probability that the given X-ray source is the real counterpart of the NIRCam source, and it can be calculated as

\begin{equation}
    R_{ij}=1-\frac{N_{gt}}{N_{sim}}
    \label{eq:Rij}
\end{equation}

where Rij is the reliability that the given \textit{ij} pair is a real coincidence, $N_{sim}$ is the number of simulated $LR$ values (200000), and $N_{gt}$ is the number of simulated $LR$ values higher than the one observed between the \textit{ij} pair $N_{gt}=N(LR_{simul} > LR_{ij})$. This reliability is then compared with a given threshold. To estimate the threshold, we merged the NIRCam catalog with a simulated X-ray catalog whose coordinates were randomly drawn from an uniform distribution, repeating the process ten times for each threshold value and taking the mean of the resulting matches. We adopted a value for the threshold equal to 0.68, which results in a number of spurious coincidences which is about 10\% that of the real matches. \par

Multiple matches have been resolved in two steps. For those cases in which multiple NIRCam sources were matching a single X-ray source, we have checked the positions of the NIRCam sources in a set of color-color and color-magnitude diagrams. If some of the NIRCam counterparts have colors typical of candidate cluster members, we have removed those compatible with being a foreground/background source. In those cases in which a single NIRCam source matched multiple X-ray sources, we have compared the median photon energy of the various X-ray sources, removing those too soft or hard for being candidate cluster members, if any. The procedure results in 2170 matched X-ray sources, over 3888 falling within the NIRCam filters, with 189 multiple matches. \par

\begin{figure}
   \centering
   \includegraphics[width=0.5\textwidth]{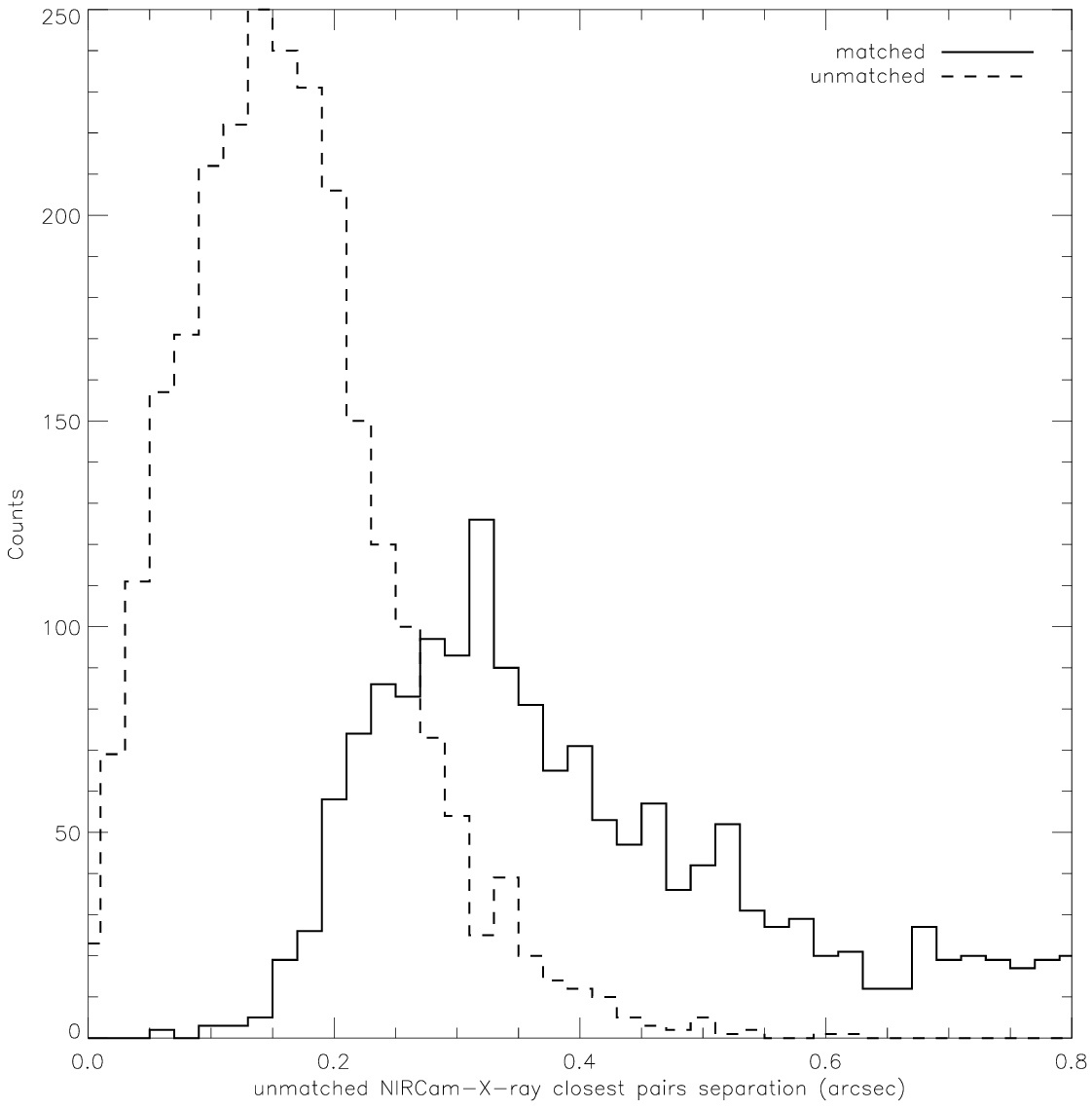}
   \caption{Distribution of the separation between the matched X-ray/NIRCam pairs (dashed histogram) and that of between the unmatched X-ray sources the closest unmatched NIRCam source.}
   \label{Fig:separations}
   \end{figure}

\begin{figure}
   \centering
   \includegraphics[width=0.5\textwidth]{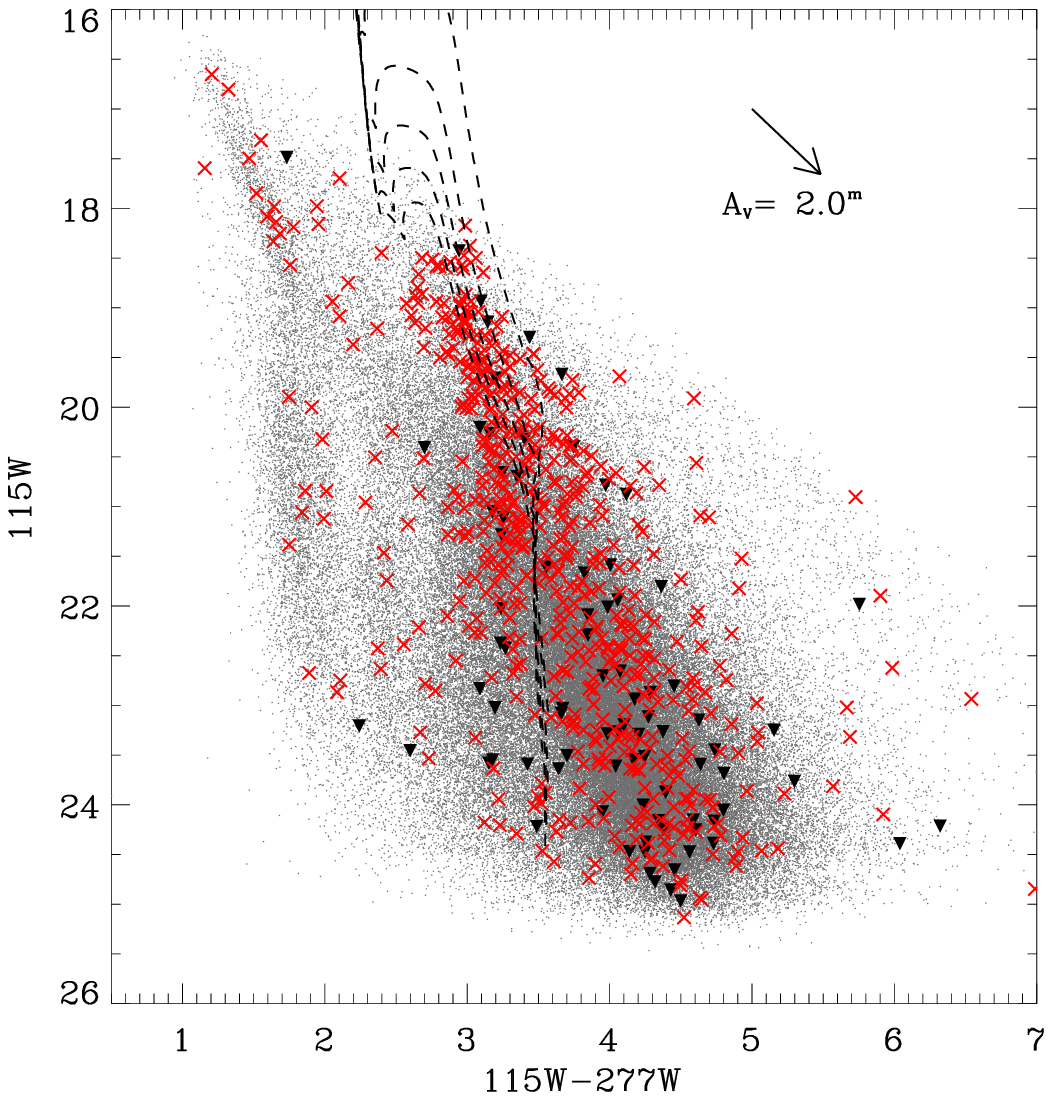}
   \caption{F115W vs. F115W-F277W diagram with marked the closest NIRCam sources to unmatched X-ray sources, in red if the separation is smaller than 0.25 arcsec, otherwise in black.}
   \label{Fig:diagramsunmatched}
   \end{figure}

\begin{figure}
   \centering
   \includegraphics[width=0.5\textwidth]{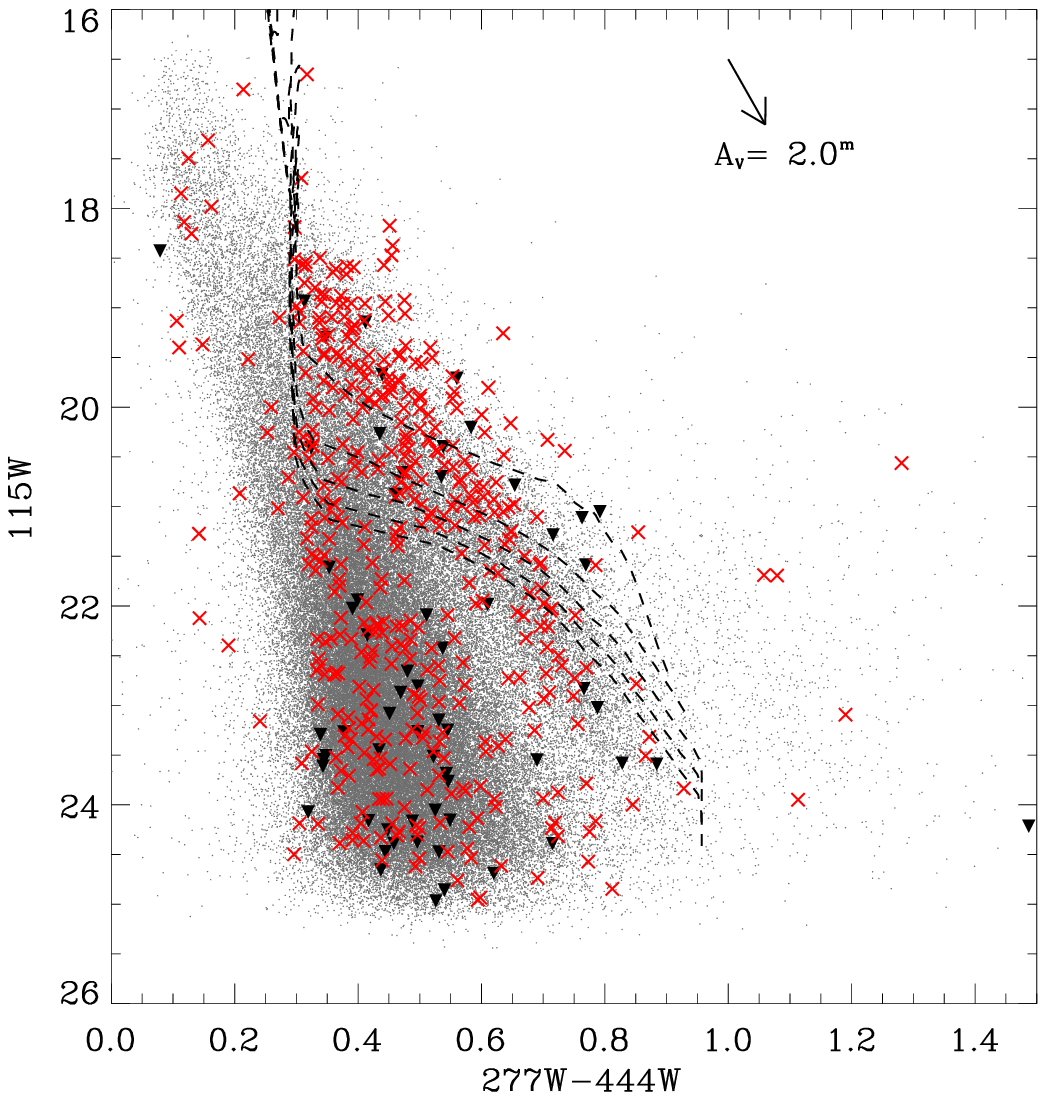}
   \caption{F115W vs. F277W-F444W diagram as Fig. \ref{Fig:diagramsunmatched}}
   \label{Fig:diagramsunmatched2}
   \end{figure}

Fig. \ref{Fig:separations} shows the distributions of the separation between X-ray and NIRCam sources both in the case of matched sources and between the unmatched X-ray sources with the closest unmatched NIRCam source. The latter distribution peaks between 0.2 and 0.4 arcsec, with only 126 pairs with a separation smaller than 0.2 arcsec. The colors of these NIRCam sources lying less than 0.2 arcsec away from unmatched X-ray sources can be seen in Figs. \ref{Fig:diagramsunmatched} and \ref{Fig:diagramsunmatched2}. They typically show colors of background sources. The two color-magnitude diagrams also show the colors of the closest NIRCam sources to unmatched sources with separations larger than 0.2 arcsec. The fact that a good fraction of these sources lie in the cluster locus does not necessarily means that they are mainly false negatives produced in the matching procedure, given that the X-ray sources mainly populate the area within and surrounding the cluster, and thus it is likely that unmatched X-ray sources have nearby cluster members that have not been detected in X-rays. \\

\begin{figure}
   \centering
   \includegraphics[width=0.5\textwidth]{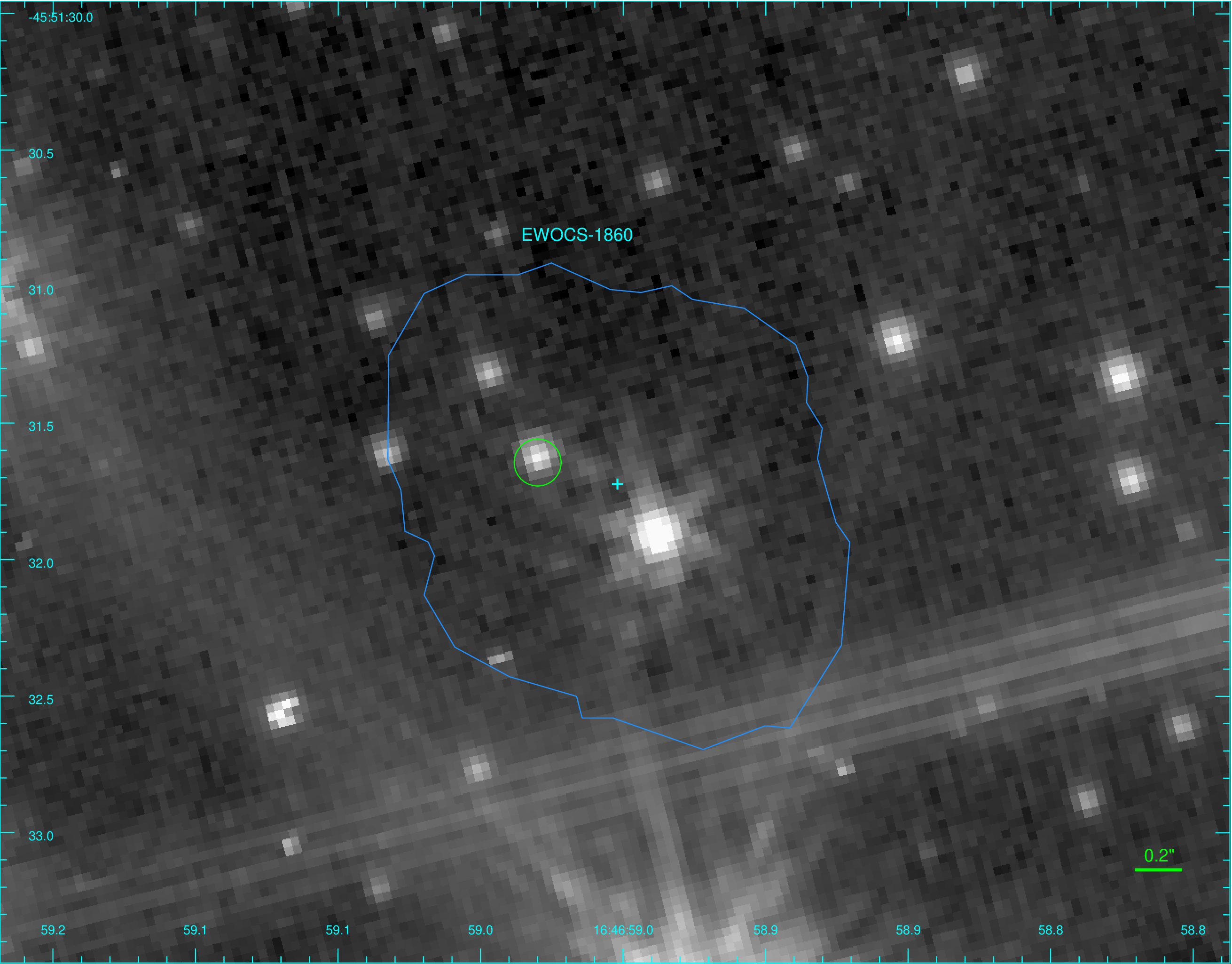}
   \caption{NIRCam image of the area surrounding the position of the EWOCS-1860 X-ray source, with two possible counterparts lying well within the X-ray extraction region (which is marked in blue): a saturated star next to the center, and the unsaturated source marked with a circle, which is likely a cluster member.}
   \label{Fig:unmatchedexample}
   \end{figure}

Fig. \ref{Fig:unmatchedexample} shows an example of a complicated determination of the true NIRCam counterpart of an X-ray sources. The EWOCS-1860 X-ray source can easily be the X-ray counterpart of the saturated star next to the center, which lies less than 0.2 arcsec from the position of the X-ray source. However, the closest unsaturated source is also likely a cluster member ([F115W-F277W]=3.062, [F115W]=19.565), lying a 0.32 arcsec away from the X-ray source, but with a corresponding reliability $R_{ij}=0.35$, much lower than our adopted threshold.   

\newpage

\end{appendix}
\end{document}